\documentclass{IEEEoj-data}
\usepackage[T1]{fontenc}
\usepackage{cite}
\usepackage{amsmath,amssymb,amsfonts}
\usepackage{graphicx,color}
\usepackage{textcomp}
\usepackage{dirtree}
\usepackage{tabularx}
\usepackage{xcolor}
\let\labelindent\relax
\usepackage{enumitem}
\usepackage{titlesec}
\usepackage{adjustbox}
\usepackage{makecell}
\usepackage{booktabs}
\usepackage[hidelinks]{hyperref}
\setlist[itemize]{noitemsep, topsep=0pt}
\titlespacing*{\subsubsection}
  {0pt}   
  {0.5ex plus 0.2ex} 
  {0.5ex plus 0.2ex}
\def\BibTeX{{\rm B\kern-.05em{\sc i\kern-.025em b}\kern-.08em
    T\kern-.1667em\lower.7ex\hbox{E}\kern-.125emX}}
\AtBeginDocument{\definecolor{ojcolor}{cmyk}{0.93,0.59,0.15,0.02}}

\makeatletter
\def\ps@preprint{%
  \def\@oddhead{}%
  \def\@evenhead{}%
  \def\@oddfoot{\hfil\normalfont\footnotesize\thepage\hfil}%
  \let\@evenfoot\@oddfoot
}
\let\ps@plain\ps@preprint
\let\ps@headings\ps@preprint
\let\ps@IEEEtitlepagestyle\ps@preprint
\makeatother
\renewcommand{\ieeelogo}{}
\hypersetup{pdftitle={Residential Electricity Consumption Dataset for Sri Lanka (RECON-SL)}}

\begin{document}
\pagestyle{preprint}

\title{\textcolor{black}{Descriptor:} \textcolor{ieeedata}{\textit{ Residential Electricity Consumption Dataset for Sri Lanka}} (RECON-SL)}

\author{CHANUKA ALGAMA\authorrefmark{1},
        MERL CHANDANA\authorrefmark{1},
        ISURUNI FERNANDO\authorrefmark{1},
        DINITHI DISSANAYAKE\authorrefmark{2},
        AMANDA ARIYARATNE\authorrefmark{1},
        NIPUNI HABARAGAMUWA\authorrefmark{1},
        PASINDU RANAGE\authorrefmark{1},
        JES\'US RAMOS\authorrefmark{1},
        and KASUN AMARASINGHE\authorrefmark{1,3}%
}
\affil{\authorrefmark{} LIRNEasia, Sri Lanka}
\affil{\authorrefmark{} National University of Singapore, Singapore}
\affil{\authorrefmark{} Carnegie Mellon University, United States}

\corresp{CORRESPONDING AUTHOR: MERL CHANDANA (e-mail: merl@lirneasia.net).}

\authornote{This work was supported by the LACUNA Fund.}

\markboth{Residential Electricity Consumption Dataset for Sri Lanka (RECON-SL)}{Chandana \textit{et al.}}

\begin{abstract}
This article describes the Residential Electricity Consumption Dataset for Sri Lanka (RECON-SL), a multi-source resource that integrates utility records, high-frequency smart meter data, and household surveys from the service area of Lanka Electricity Company (LECO) in Sri Lanka. The dataset captures electricity use for 4,063 households, collected over the period October 2022 to January 2025, with coverage at monthly, 6-hour, and 15-minute intervals. Monthly billing-cycle data provide complete coverage across all households, while smart meter data are available for a subsample of 1,438 households, offering both fine-grained 15-minute readings and coarser 6-hour records. Three survey waves complement these consumption measures, collecting information on demographics, housing, appliance ownership, and perceptions of energy security.\\
\\
Across its components, the dataset includes more than 50 million meter readings and detailed survey responses from over 11,000 household interviews. As the first dataset of its kind in Sri Lanka, and among the few worldwide to link smart meter records with longitudinal survey data at this scale, it provides a unique resource for energy,  machine learning and policy research, with known gaps in smart meter coverage documented for users. The dataset can support diverse applications, including electricity load forecasting, socio-economic and behavioral energy research, distribution planning, and policy analysis in energy affordability and equity. Access is provided through an open repository, with accompanying documentation and scripts to facilitate reuse.\\
 \\
 {\textcolor{ieeedata}{\abstractheadfont\bfseries{IEEE SOCIETY/COUNCIL}}}     Power and Energy Society (PES)\\
 \\
 {\textcolor{ieeedata}{\abstractheadfont\bfseries{DATA DOI/PID}}} \href{https://doi.org/10.21227/n1dk-q860}{10.21227/n1dk-q860} \\

 {\textcolor{ieeedata}{\abstractheadfont\bfseries{DATA TYPE/LOCATION}}}  Time-series; Survey; Longitudinal residential electricity consumption and behavior data; Sri Lanka

\end{abstract}


\maketitle

\raggedbottom

\section*{BACKGROUND}

\begin{figure}[!t]
\centering
\includegraphics[width=\columnwidth]{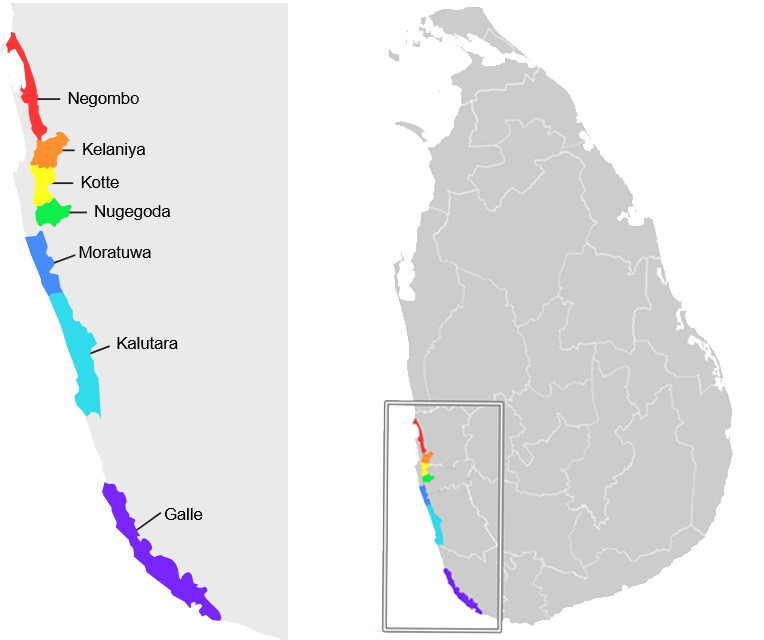}
\caption{Geographical coverage of the Lanka Electricity Company (LECO) operational areas sampled for the RECON-SL dataset.
LECO supplies electricity to about 500,000 households across seven coastal branches---Negombo, Kelaniya, Kotte, Nugegoda, Moratuwa, Kalutara, and Galle---shown here in color.}
\label{fig:map}
\end{figure}

The dataset introduced in this paper consists of two parts:  electricity consumption data and longitudinal survey data. The electricity consumption data span over 20 months, while the longitudinal survey data comprise three rounds of surveys---referred to as Wave 1, Wave 2, and Wave 3---conducted over 15 months across more than 4,000 households in Sri Lanka. These households were selected from the customer base of Lanka Electricity Company (LECO), one of the island's two main electricity distributors. LECO serves approximately 500,000 households across 7 regional branches, primarily along the western coastal belt of Sri Lanka, as illustrated in Figure~\ref{fig:map}. Among these households, about 10\% are equipped with smart meters, which record electricity consumption at 15-minute intervals throughout the day. The remaining households use conventional meters that provide monthly readings of their consumption. Given the relatively low penetration of smart meters in Sri Lanka's residential sector, this dataset presents a valuable opportunity to analyze high-frequency electricity consumption patterns in conjunction with household-level survey data, thereby simulating broader smart-meter readings, building predictive models, and validating subsidy and tariff structures.

\subsection*{Dataset Composition}

\subsubsection*{Electricity Consumption Data}

\textbf{Monthly consumption data:} Monthly consumption values are available for all participating households over the study period. These records capture cumulative electricity usage at the end of each billing cycle, providing a consistent measure of household demand.\par
\textbf{Smart meter data(15-minute interval):} A subset of households---approximately 10\% of LECO's overall customer base, corresponding to about 35\% of the households in this dataset---is equipped with smart meters that record electricity consumption at 15-minute intervals. These high-frequency readings capture daily load profiles and temporal variability, enabling detailed analysis of peak demand, diurnal usage patterns, and household consumption dynamics. Missing intervals and outages are documented to guide appropriate use.\par
\textbf{Smart meter data (6-hour):} In addition to the 15-minute interval records, LECO maintains a complementary dataset at 6-hour resolution. Although derived from the same smart meters, these data are processed and stored separately by the utility as part of routine monitoring and quality assurance operations. The 6-hour files contain some features not available in the raw 15-minute stream and were used to address observed gaps in the higher-frequency data. Aggregated values are aligned with household identifiers and survey waves, facilitating integration with the broader dataset.\par

\subsubsection*{Survey Data}

\textbf{Wave 1 (baseline survey):} The first wave established a comprehensive baseline across more than 4,000 households. Core modules included demographics, housing and appliance stock, room and energy system characteristics, and general household information. These provide the foundation for linking household context with electricity consumption and set the reference point for subsequent waves.\par
\textbf{Wave 2 (follow-up survey):} The second wave revisited the panel to capture change over time, focusing on appliance turnover, modifications to rooms, and behavioral adjustments in electricity use. It repeated several baseline modules to allow comparability and added targeted items such as conscious consumption reduction. This wave also documented residents' energy usage practices and attitudes towards energy-saving.\par
\textbf{Wave 3 (final survey):} The third wave extended the panel with emphasis on affordability, tariff perceptions, and energy security, while continuing to track appliance and room changes. New modules captured payment practices, awareness of time of use billing, and conscious reductions in appliance usage alongside continuity in appliance ownership and occupancy.

\begin{figure}[!b]
    \centering
    \includegraphics[width=\columnwidth]{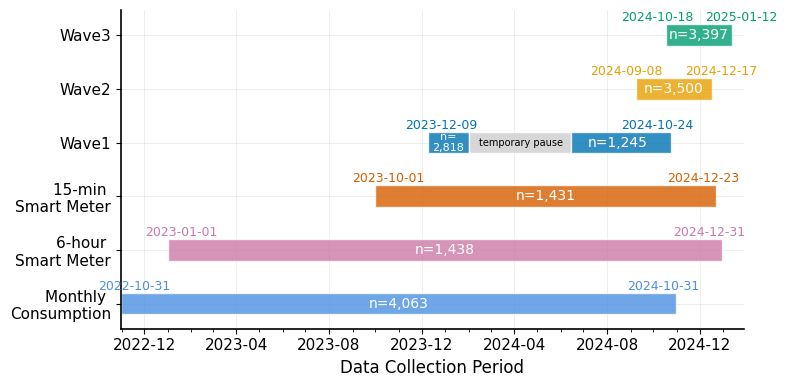}
    \caption{Data collection timeline for the RECON-SL dataset. The sample sizes for each component are annotated ($n$ values), along with start and end dates, highlighting periods of overlap and interruptions.}
    \label{fig:timeline}
\end{figure}

\begin{figure*}[t]
    \centering
    \includegraphics[width=16cm]{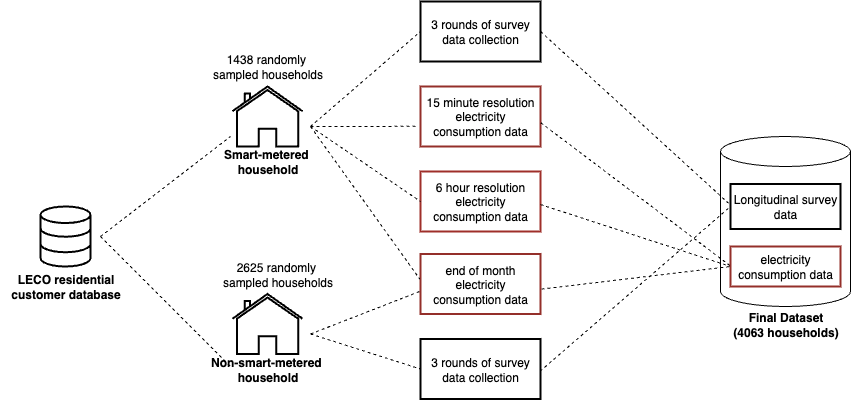}
    \caption{Composition of the RECON-SL dataset, containing 1,438 smart-metered and 2,625 non-smart-metered households, contributing electricity consumption data at 15-minute, 6-hour, and monthly resolutions. All 4,063 households participated in at least the first round of surveys, producing a combined dataset of longitudinal household information and electricity use.}
    \label{fig:data-collection}
\end{figure*}

Figure~\ref{fig:timeline} illustrates the data collection timeline across different sub-components of the dataset, while Figure~\ref{fig:data-collection} provides a visual snapshot of the final dataset.

\subsection*{Comparable Published Datasets}

A growing number of openly available residential energy datasets have shaped the field in recent years. These resources typically fall into two broad families. Some focus on fine-grained metering---smart meter or submeter recordings at resolutions ranging from seconds to half-hours---often collected from small samples but with high temporal resolution. Others emphasize survey and contextual information, covering demographics, housing, appliance ownership, and expenditures ---frequently at large scale but without linked high-frequency load data. With a few notable exceptions, the majority of these dataset is from Europe and North America, where infrastructure, data governance, and research investment enable long-term data capture. Together, these datasets have provided the benchmarks for algorithm development in non-intrusive load monitoring, forecasting, and demand-side management. Few noteworthy datasets include:

\begin{enumerate}
    \item UK-DALE (UK): Whole-home and appliance-level electricity demand for five homes (up to 16 kHz/1-s), widely used for NILM and load-shape research. \cite{DALE2015}.
    \item IDEAL Household Energy Dataset (UK): 255 homes over ~23 months; electricity (1-s), gas, room sensors, and contextual/survey data---rich multimodal signals for residential analytics. \cite{IDEAL2021}
    \item REFIT (UK): Two-year longitudinal measurements (8-s) from 20 homes at aggregate and appliance level---frequently used for forecasting and NILM benchmarking \cite{REFIT2017}.
    \item SustDataED2 (Portugal): A residential labeled dataset for smart meter data analytics includes labeled appliance usage and aggregated data; useful for learning appliance ON/OFF transitions \cite{pereira2022residential}
    \item RECS (USA): Nationally representative residential energy consumption survey (household characteristics, fuels, expenditures)---an anchor for socio-economic analyses though not linked to high-frequency meters.\cite{EIA_RECS2020}
    \item iAWE (India): An Indian residential dataset with meter, circuit, and appliance-level measurements---a key Global South benchmark for NILM/behavior studies \cite{batra2013s}
    \item PRECON (Pakistan): residential electricity consumption for ~42 properties, varied demographics, over one year; includes whole-house and high-powered device readings \cite{nadeem2019precon}

\end{enumerate}

\subsection*{RECON-SL's Positioning Relative to Existing Datasets}

Relative to these resources, the present dataset's distinctive features are: (i) multi-resolution utility data (monthly + 6-hr + 15-min) at thousand-home scale, (ii) linkage to three survey waves covering demographics, housing, appliance stock, and perceptions, and (iii) location in a South Asian, emerging-economy context. This combination is uncommon: high-frequency smart-meter corpora rarely include longitudinal surveys at scale; survey-heavy assets lack linked high-frequency load; city-scale streams typically omit rich household attributes.

The dataset enables load modeling across temporal scales (15-min/6-hr/monthly), behavioral linkage (appliance ownership, household composition, attitudes), prosumer/solar export analyses where applicable, and policy-relevant evaluation (e.g., tariff changes) using a longitudinal survey and smart meter data --- capabilities that are rarely co-present in open datasets from the Global South. It complements existing consumption datasets, while adding regional diversity and multi-wave behavioral context for researchers and policymakers focusing on equity, affordability, and demand-side interventions.

\subsection*{Suitable and Potential Use Cases}
This dataset enables a range of impactful research applications:

\textbf{Machine learning \& forecasting:} Load forecasting across temporal scales (15-min to monthly) using federated and centralized models \cite{fekri2022distributed}; appliance-ownership prediction via combined smart meter and survey data \cite{saraf2024appliance}; consumption clustering for demand response modeling \cite{khan2019smart}\par
\textbf{Behavioral \& socio-economic analysis}: Integration of survey and meter data to profile energy behaviors \cite{adams2021smart}; longitudinal meter--survey linkage to study tariff reforms or behavioral adaptations \cite{tang2022machine}\par
\textbf{Energy systems research:} Creating distribution-level load profiles with smart meter data \cite{dewangan2023load}; analyzing prosumer solar export dynamics and net-metering impacts \cite{khan2019smart}\par
\textbf{Policy \& social science:} Assessing equity and affordability across socio-economic groups; exploring gendered usage patterns (context permitting); linking perceived energy security to actual reliability metrics \cite{adams2021smart}\par

\subsection*{Prior uses of this dataset:}
The dataset has been made publicly available through IEEE DataPort, Zenodo, and Kaggle, where it has been downloaded more than 200 times at the time of writing. No peer-reviewed publications using this dataset have been identified to date. Ongoing work by the authors includes a study applying machine learning techniques to identify households with inefficient electricity consumption patterns.

\section*{COLLECTION METHODS AND DESIGN}

\subsection*{Study Population and Sampling}

Households were drawn from the customer base of the Lanka Electricity Company (LECO), which distributes electricity to ~500,000 households along Sri Lanka's western coastal belt. A stratified random sampling design was used, balancing smart-metered and conventional-metered households and ensuring coverage across LECO branch areas. The original target sample was 4,000 households; during implementation, 4,063 participated in Wave 1, 3,500 in Wave 2, and 3,397 in Wave 3, with an overall retention rate above 80\%. Replacement households were recruited from the same strata when initial participants could not be reached after repeated attempts.

\subsection*{Electricity Consumption Data Collection}

Monthly billing data were available for all surveyed households over a 20-month period, derived from LECO's operational meter-reading systems. A subset of households equipped with smart meters recorded consumption at 15-minute intervals using LECO's existing infrastructure, with no additional instrumentation required. To address gaps in the high-frequency stream caused by equipment or server malfunctions, LECO also provided a complementary 6-hour dataset, generated and stored separately as part of its routine monitoring operations. All consumption data were shared under a formal agreement with LECO and linked to household identifiers for integration with the survey component.

\subsection*{Survey Data Collection}

Three longitudinal survey waves were conducted over a 15-month period. The baseline survey (Wave 1) captured household room rosters, demographics, housing characteristics, appliance stock, energy use behaviors, and solar adoption. The follow-up (Wave 2) documented appliance purchases and disposals, changes in usage behaviors such as laundry, ironing, and cooking, and evolving attitudes toward energy services. The final survey (Wave 3) focused on billing and payment practices, device-use routines, awareness of time-of-use tariffs, and continuity checks. All surveys were administered through computer-assisted personal interviewing (CAPI) on tablets, translated into Sinhala and Tamil, and piloted for comprehension and timing. Respondents were typically household heads or knowledgeable proxies, with efforts made to maintain respondent consistency across waves.\par Table \ref{tab:survey-summary} provides a summary of modules across the waves while a comprehensive account of the survey design and administration is provided in the Survey Methodology Note (\hyperref[app:A]{Appendix A}).

\begin{table*}[!htbp]
\scriptsize
\centering
\caption{Survey module availability across three longitudinal waves of data collection.}
\label{tab:survey-summary}
\begin{tabularx}{\textwidth}{p{3cm} p{4.8cm} p{4.8cm} ccc}
\hline
\textbf{Module Name}\rule{0pt}{2.6ex} & \textbf{Description} & \textbf{Key Data} & \multicolumn{3}{c}{\textbf{Available in}} \\
\cline{4-6}
 & & & \textbf{Wave 1}\rule{0pt}{2.6ex} & \textbf{Wave 2} & \textbf{Wave 3} \\
\hline
ac\_roster\rule{0pt}{2.6ex} & Specifications of air conditioning units and usage hours & type, BTU, no of hours used in the previous week & \checkmark & \checkmark & \checkmark \\
appliances, appliance\_usage & Household appliances and usage hours & type, no of hours used in the previous week & \checkmark & \checkmark & \checkmark \\
awareness\_of\_time\_of\_use & Household characteristics related to time-of-use electricity pricing & use of ToU metering, if consumption reduced, ability to shift consumption patterns & \texttimes & \texttimes & \checkmark \\
behaviour & General energy consumption behavioural patterns & behaviour around ironing clothes, drying clothes, lighting, refrigerator use & \texttimes & \checkmark & \texttimes \\
behaviour\_data & Behaviour about consciously reducing electricity consumption & type of appliances of which consumption was reduced consciously & \texttimes & \texttimes & \checkmark \\
changed\_appliances & Changes in household appliances between survey waves & type and no of appliances changed & \texttimes & \checkmark & \checkmark \\
changed\_rooms & Modifications to room structures between survey waves & type of rooms, no of new rooms since last survey  & \texttimes & \checkmark & \checkmark \\
demographics & Household demographic information (age, education, income, etc.) & age, education, occupation of house members, no of hours spent at home during the previous week & \checkmark & \texttimes & \texttimes \\
demolished\_rooms & Rooms that were removed between survey waves & type of room, no of rooms removed & \texttimes & \texttimes & \checkmark \\
electricity\_billing\_info & Electricity billing and payment information & target units, target bill value, payment practices & \texttimes & \texttimes & \checkmark \\
electricity\_generation & Changes to solar energy system since previous survey & new installations, new panels added or removed, system details & \texttimes & \checkmark & \checkmark \\
electricity\_generation\_water \_heating\_cooking & Energy sources for electricity generation, water heating, and cooking & energy generation methods, solar system specs, water heating, water heating and cooking methods & \checkmark & \texttimes & \texttimes \\
fan\_roster & Usage patterns of household fans & type, no of hours used in the previous week & \checkmark & \checkmark & \checkmark \\
household\_information & General household characteristics and information & details of renters, renovations, architecture decisions & \texttimes & \checkmark & \checkmark \\
household\_information\_and \_history & Household information with historical context & city, built year, building materials, no of floors, floor area, socio-economic class & \checkmark & \texttimes & \texttimes \\
light\_roster & Usage patterns of light bulbs & type, wattage, no of hours used in the previous week & \checkmark & \checkmark & \checkmark \\
members\_who\_were\_in\_w1 & Tracking of household members from Wave 1 & new members, members who left, changed occupations, no of hours stayed home in the previous week & \texttimes & \checkmark & \texttimes \\
members\_who\_were\_in\_w2 & Tracking of household members from Wave 2 & new members, members who left, changed occupations, no of hours stayed home in the previous week & \texttimes & \texttimes & \checkmark \\
new\_household\_members & New members added to household between waves & age, education, occupation of new members & \texttimes & \checkmark & \checkmark \\
room\_roster & Specifications of the rooms within the households & type of room, building material, no of doors, windows, bulbs, ACs & \checkmark & \checkmark & \checkmark \\
\hline
\end{tabularx}
\end{table*}

\section*{VALIDATION AND QUALITY}

\subsection*{Completeness of Meter Data}
Monthly billing data are complete across the full sample, whereas smart meter data show more variable coverage, with gaps arising from communication failures, power outages, and server malfunctions. Of the 1,788 households surveyed with smart meters in Wave 1, valid consumption records were retained for 1,438; the remainder are effectively non smart-metered in the final dataset. For the full intended observation period, overall record completeness is 20.2\% for the 15-minute dataset and 47.7\% for the 6-hour dataset. When restricting to the windows where records are available, completeness rises to 35.5\% and 51.7\%, respectively.\par
Beyond aggregate completeness percentages, we also report daily and weekly coverage thresholds for the smart meter subsets (Table ~\ref{tab:completeness-summary}). These indicators show that while overall completeness is 20.2\% for 15-minute data and 47.7\% for 6-hour data, coverage is more robust at coarser temporal units: for example, 89.3\% of days in the 6-hour dataset have records from at least half of households. These metrics are meant to provide a clearer sense of the dataset's practical usability for load profiling and forecasting applications.\par
\begin{figure*}[!tp]
\centering
\includegraphics[width=\textwidth]{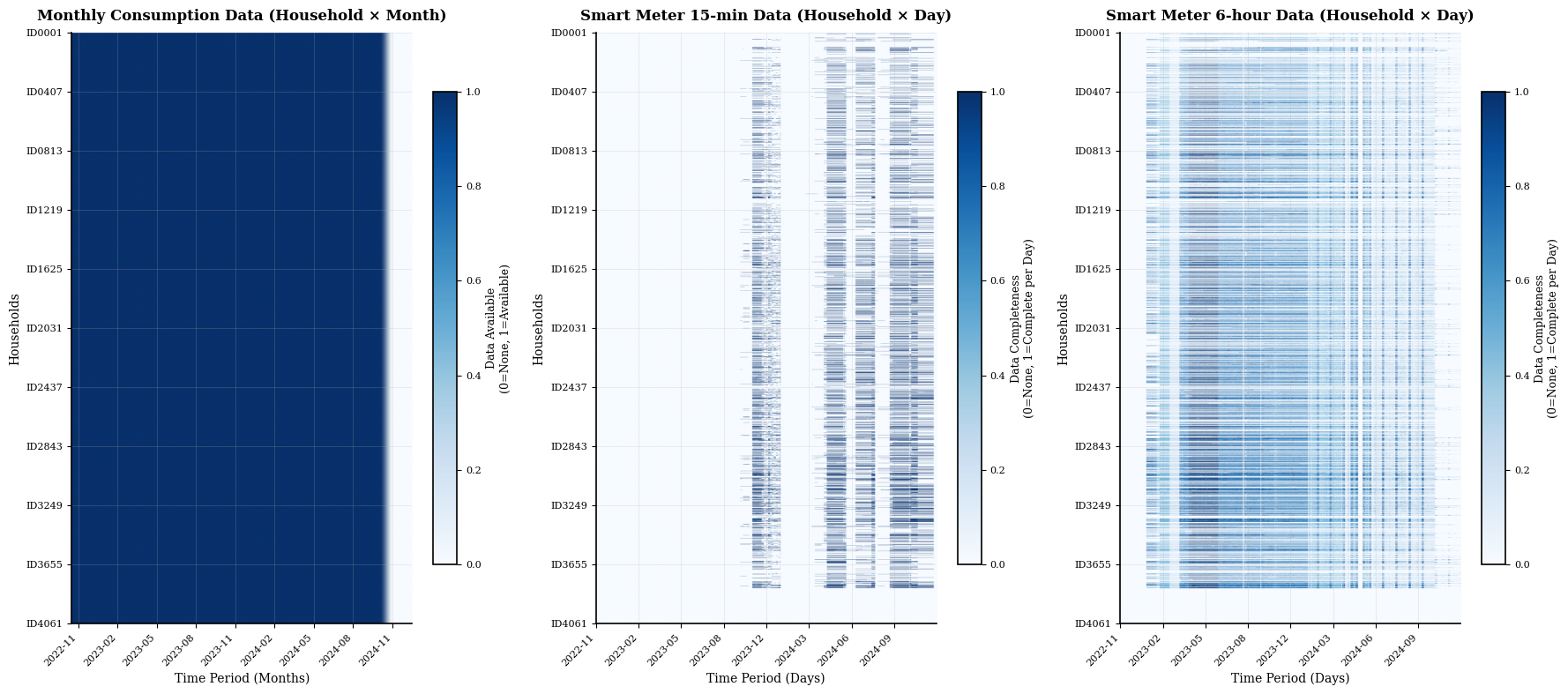}
\caption{Coverage of electricity consumption data across households and time.
(Left) Monthly billing data (\texttt{household} \texttimes{} \texttt{month}),
(Middle) smart meter 15-minute data (\texttt{household} \texttimes{} \texttt{day}),
(Right) smart meter 6-hour data (\texttt{household} \texttimes{} \texttt{day}).
Darker shading indicates more complete records, while lighter shading indicates missing values or partial coverage.}
\label{fig:consumption-coverage}
\end{figure*}

 Household-level heatmaps (Figure \ref{fig:consumption-coverage}) illustrate missingness patterns, distinguishing extended contiguous gaps.\par

\begin{table*}[!htbp]
\centering
\small
\caption{Data completeness across monthly meter readings, smart meter data (15-min and 6-hour), and three survey waves. The table summarizes household coverage, record counts, and completeness relative to both requested and available date ranges, highlighting variations in temporal coverage.}
\label{tab:completeness-summary}
\begin{adjustbox}{max width=\linewidth}
\begin{tabular}{l*{8}{r}}
\toprule
Component &
\thead{Monthly} &
\thead{Smart Meter\\15min} &
\thead{Smart Meter\\6hour} &
\thead{Survey\\Wave1} &
\thead{Survey\\Wave 2} &
\thead{Survey\\Wave 3} \\
\midrule
Households Intended & 4,000 & 2,000 & 2,000 & 4,000 & 4,000 & 4,000 \\
Households with Data & 4,063 & 1,431 & 1,438 & 4,063 & 3,500 & 3,397 \\
Expected Records for Requested Date Range & 109,701 & 108,801,792 & 4,555,584 & 4,063 & 3,500 & 3,397 \\
Expected Records for Available Date Range & 101,575 & 61,819,200 & 4,204,712 & 4,063 & 3,500 & 3,397 \\
Actual Records & 101,575 & 21,970,542 & 2,174,877 & 4,063 & 3,500 & 3,397 \\
\% Completeness with Requested Dates & 92.6\% & 20.2\% & 47.7\% & 100.0\% & 100.0\% & 100.0\% \\
\% Completeness with Available Dates & 100.0\% & 35.5\% & 51.7\% & 100.0\% & 100.0\% & 100.0\% \\
\% of Days with 50\% of households having $\geq$1 record & N/A & 42.7\% & 89.3\% & N/A & N/A & N/A  \\
\% of Days with 75\% of households having $\geq$1 record & N/A & 8.2\% & 82.5\% & N/A & N/A & N/A \\
\% of Weeks with at least 50\% of Records for $\geq$50\% Households & N/A & 43.1\% & 63.2\% & N/A & N/A & N/A \\
\% of Weeks with at least 50\% of Records for $\geq$75\% of Households & N/A & 4.6\% & 50.9\% & N/A & N/A & N/A \\
\bottomrule
\end{tabular}
\end{adjustbox}
\end{table*}

\subsection*{Consistency and Internal Validation}
Monthly aggregates from smart meter households were compared against corresponding LECO billing records to verify that the high-frequency streams were consistent with the utility's official billing data. This step ensured that smart meter readings reliably captured household consumption and provided confidence for subsequent integration with survey variables. Discrepancies were within an acceptable range. The 6-hour dataset was validated against the underlying 15-minute readings to confirm agreement in cumulative consumption and to further establish the fidelity of both smart meter data sources. Histograms and diurnal load profiles were compared across smart- and non-smart-meter households to assess plausibility of consumption patterns. These confirmed that consumption values were within reasonable ranges and temporal patterns reflected expected daily cycles, within-week variation, and seasonal shifts.

\subsection*{Survey Data Quality}
Attrition across survey waves was relatively low, with retention reaching 83.6\%. Most attrition arose either from household migration or from the inability to reach respondents within the survey window for follow-up interviews.\par
Multiple verifications were built in to the data collection process, including accompaniments where supervisors observed interviews in real time, back-checks covering approximately 18\% of households, and spot checks that verified around 10\% of interviews on site. In addition, telephonic audits reached about 80\% of households to confirm key items, GPS verification was carried out for 100\% of interviews to confirm reported locations, and partial voice audits were conducted for roughly 60\% of cases to ensure quality.\par
Further, logic checks were applied across survey waves. Skip patterns in the instrument were tested, appliance counts were cross-checked over waves, and responses were reviewed against commonsense thresholds to identify implausible entries. Demographic variables were monitored for stability. The full survey methodology is provided in \hyperref[app:A]{Appendix A}, and the complete questionnaire is included in \hyperref[app:B]{Appendix B}.

\subsection*{Known Limitations}
The dataset has several limitations. Only about 10\% of the LECO customer base contributes smart meter data, and is primarily a result of the smart-meter roll-out plan by LECO. Geographic coverage is also restricted to LECO's service areas, which is concentrated in urban and western coastal regions. Residual missingness in meter records---particularly contiguous gaps, reduces the value of the data for fine-grained load disaggregation. Certain self-reported survey variables --- such as previous month's expenditure --- may be subject to recall bias or estimation errors.

\section*{RECORDS AND STORAGE}

The dataset is organized in a hierarchical structure with two main data categories: electricity consumption data and survey data. The data is stored in multiple CSV files organized across multiple directories, based on the data collection method and temporal resolution. The complete file organization is presented in the directory tree below.

\dirtree{%
.1 data/.
.2 consumption\_data/.
.3 non\_smart\_meter/.
.4 monthly\_consumption.csv.
.3 smart\_meter/.
.4 6hour\_interval/.
.5 smart\_6hour\_[1-5].csv.
.4 15minute\_interval/.
.5 smart\_15min\_[1-3].csv.
.2 survey\_data/.
.3 wave\_1/....
.3 wave\_2/....
.3 wave\_3/....
.3 survey\_dates.csv.
.2 readme.md.
}

\begin{figure*}[!tp]
\centering
\includegraphics[width=0.95\textwidth]{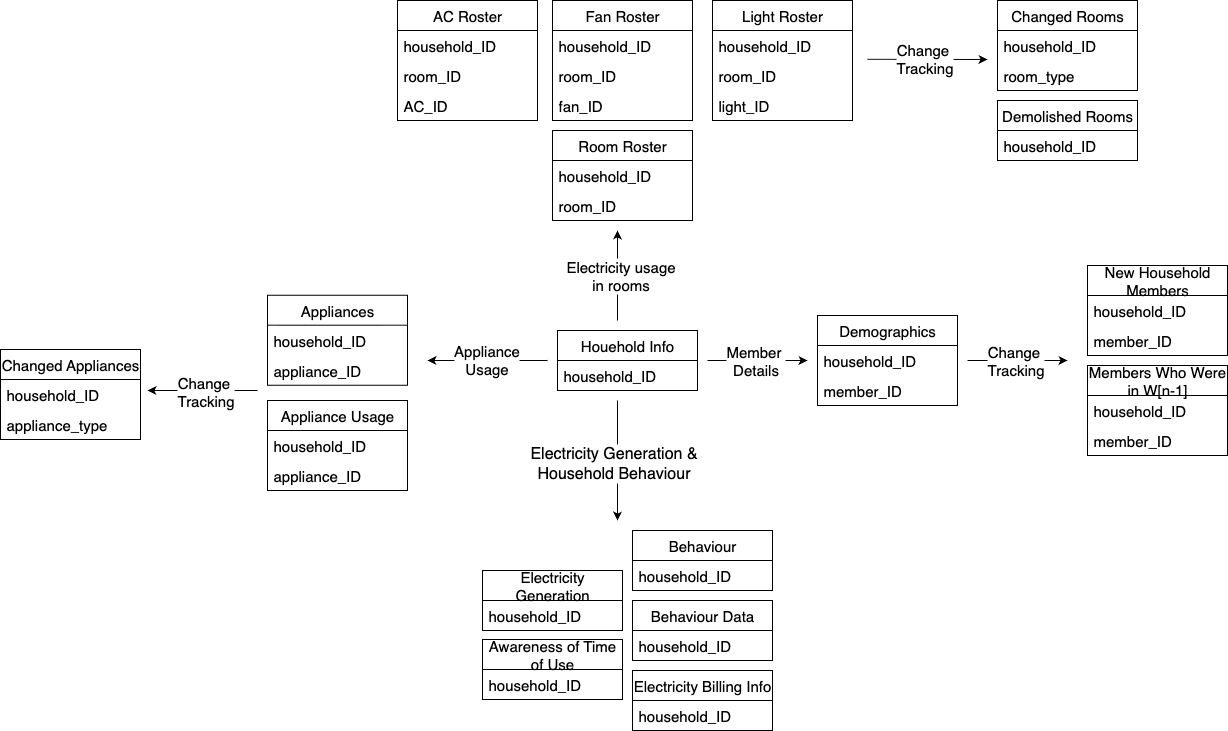}
\caption{Linking of household survey modules in the RECON-SL dataset. Core identifiers include \texttt{household\_ID} (all modules), \texttt{room\_ID} (room and appliance rosters), \texttt{appliance\_ID} (appliance usage), and \texttt{member\_ID} (demographic and member tracking). These keys allow integration of information across modules and waves.}
\label{fig:survey-linking}
\end{figure*}

Households are linked across files through a consistent household ID, combining consumption patterns with household characteristics while maintaining anonymity. Linkage is further illustrated in Figure~\ref{fig:survey-linking}. Monthly consumption data provides direct usage in kilowatt-hours and smart meter data contains cumulative readings that require calculating the differences between consecutive readings to derive consumption. The smart meter data also include households that were not part of the survey. These households were retained in the dataset for their potential standalone value, and can be separated by household IDs and don't contribute to any information presented in this paper. The survey data organization is depicted in Table~\ref{tab:survey-summary}.

\section*{INSIGHTS AND NOTES}

\subsection*{Practical Insights for Use}

Unlike appliance- or sub-metered datasets, RECON-SL is based on smart meters installed and operated by the utility during regular operations, without the need for additional measurement equipment. The linked survey provides a wide range of variables, offering multiple avenues for exploration. However, from a policy and implementation perspective, it is unlikely that such detailed surveys will be routinely conducted at scale.

 Researchers are encouraged to explore use cases that rely on a limited set of easily collectable household variables in combination with smart meter data, as these approaches are more likely to be scalable, transferable, and policy relevant.

\subsection*{Unsuitable and Cautionary Use Cases}

\textbf{National representativeness}: The sample covers only households in LECO's service area (primarily the western coastal region of Sri Lanka). Any result should not be generalized beyond LECO households within these regions.\par
\textbf{Appliance-level disaggregation}: All consumption data being at the household level cannot directly support appliance-level load disaggregation without additional assumptions.\par
\textbf{Income analysis}: The dataset does not include precise household income; proxies include previous month's expenditure and socio-economic classifications. These should not be used in analyses requiring accurate income estimates.\par
\textbf{Real-time operational forecasting}: Data are historical and subject to transmission and storage gaps; they are not suitable for live grid monitoring or operational control.\par

\subsection*{Ethics \& Anonymization}

Survey data were collected with informed consent under ethics approval obtained from the Ethics Review Committee at the University of Moratuwa. Participants were notified that their household electricity consumption data, provided by LECO, would be linked to survey responses for research purposes, with a mechanism available to withdraw consent at any time. All personally identifying details (names, addresses) have been removed, with household IDs pseudonymized and geographic information aggregated to divisional secretariat level. These measures minimize risks, re-identification of households may still be possible if the dataset is combined with external datasets. A full Datasheet for the dataset \cite{gebru2021datasheets} is provided in \hyperref[app:C]{Appendix C}, documenting ethical safeguards and anonymization procedures.

\section*{SOURCE CODE AND SCRIPTS}

The complete data processing pipeline and exploratory data analysis scripts are publicly available in the LIRNEasia Lacuna repository at \url{https://github.com/LIRNEasia/lacuna}. This repository contains all the code used for data cleaning, pre-processing, and initial analysis of both the electricity consumption and survey datasets.

\section*{ACKNOWLEDGEMENTS AND INTERESTS}
This work was funded by the LACUNA Fund as part of their awards for climate datasets in health and energy (2022 Climate Awardees). Grantee \#55 (19497.59).
\\
\\
Special thanks to the Chairman and staff of LECO for providing data access and supporting the survey.
\\
\\
M.C. was the main author and supervised all aspects of dataset development. C.A. contributed to questionnaire design and consumption data preparation. I.F. oversaw survey administration and data quality. D.D. developed the questionnaire with input from others. A.A. created diagrams and drafted paper sections. N.H. led survey data cleaning. P.R. supported data quality assurance and cleaning. J.R. assisted in data preparation. K.A. provided overall supervision and reviewed the manuscript.
\\
\\
The authors acknowledge the support of LIRNEasia staff in survey design and coordination, and Survey Research Lanka (SRL) for field implementation.
\\
\\
The article authors have declared no conflicts of interest.
\bibliographystyle{IEEEtran}
\bibliography{refs}

@article{DALE2015,
  title={The {UK-DALE} dataset, domestic appliance-level electricity demand and whole-house demand from five {UK} homes},
  author={Kelly, Jack and Knottenbelt, William},
  journal={Scientific data},
  volume={2},
  number={1},
  pages={1--14},
  year={2015},
  publisher={Nature Publishing Group}
}

@article{IDEAL2021,
  title={The IDEAL household energy dataset, electricity, gas, contextual sensor data and survey data for 255 UK homes},
  author={Pullinger, Martin and Kilgour, Jonathan and Goddard, Nigel and Berliner, Niklas and Webb, Lynda and Dzikovska, Myroslava and Lovell, Heather and Mann, Janek and Sutton, Charles and Webb, Janette and others},
  journal={Scientific Data},
  volume={8},
  number={1},
  pages={146},
  year={2021},
  publisher={Nature Publishing Group UK London}
}

@article{REFIT2017,
  title={An electrical load measurements dataset of United Kingdom households from a two-year longitudinal study},
  author={Murray, David and Stankovic, Lina and Stankovic, Vladimir},
  journal={Scientific data},
  volume={4},
  number={1},
  pages={1--12},
  year={2017},
  publisher={Nature Publishing Group}
}

@article{fekri2022distributed,
  title={Distributed load forecasting using smart meter data: Federated learning with Recurrent Neural Networks},
  author={Fekri, Mohammad Navid and Grolinger, Katarina and Mir, Syed},
  journal={International Journal of Electrical Power \& Energy Systems},
  volume={137},
  pages={107669},
  year={2022},
  publisher={Elsevier}
}

@article{khan2019smart,
  title={Smart meter data based load forecasting and demand side management in distribution networks with embedded PV systems},
  author={Khan, Zafar A and Jayaweera, Dilan},
  journal={IEEE Access},
  volume={8},
  pages={2631--2644},
  year={2019},
  publisher={IEEE}
}

@article{adams2021smart,
  title={How smart meter data analysis can support understanding the impact of occupant behavior on building energy performance: A comprehensive review},
  author={Adams, Jacqueline Nicole and Belafi, Zsofia Deme and Horv{\'a}th, Mikl{\'o}s and Kocsis, J{\'a}nos Bal{\'a}zs and Csoknyai, Tam{\'a}s},
  journal={Energies},
  volume={14},
  number={9},
  pages={2502},
  year={2021},
  publisher={MDPI}
}

@article{tang2022machine,
  title={Machine learning approach to uncovering residential energy consumption patterns based on socioeconomic and smart meter data},
  author={Tang, Wenjun and Wang, Hao and Lee, Xian-Long and Yang, Hong-Tzer},
  journal={Energy},
  volume={240},
  pages={122500},
  year={2022},
  publisher={Elsevier}
}

@article{dewangan2023load,
  title={Load forecasting models in smart grid using smart meter information: A review},
  author={Dewangan, Fanidhar and Abdelaziz, Almoataz Y and Biswal, Monalisa},
  journal={Energies},
  volume={16},
  number={3},
  pages={1404},
  year={2023},
  publisher={MDPI}
}

@article{pereira2022residential,
  title={A residential labeled dataset for smart meter data analytics},
  author={Pereira, Lucas and Costa, Donovan and Ribeiro, Miguel},
  journal={Scientific Data},
  volume={9},
  number={1},
  pages={134},
  year={2022},
  publisher={Nature Publishing Group UK London}
}

@article{gebru2021datasheets,
  title={Datasheets for datasets},
  author={Gebru, Timnit and Morgenstern, Jamie and Vecchione, Briana and Vaughan, Jennifer Wortman and Wallach, Hanna and Iii, Hal Daum{\'e} and Crawford, Kate},
  journal={Communications of the ACM},
  volume={64},
  number={12},
  pages={86--92},
  year={2021},
  publisher={ACM New York, NY, USA}
}

@inproceedings{saraf2024appliance,
  title={Appliance ownership prediction with smart meter data},
  author={Saraf, Anmol and Kowli, Anupama},
  booktitle={Proceedings of the 15th ACM International Conference on Future and Sustainable Energy Systems},
  pages={633--638},
  year={2024}
}

@inproceedings{nadeem2019precon,
  title={PRECON: Pakistan residential electricity consumption dataset},
  author={Nadeem, Ahmad and Arshad, Naveed},
  booktitle={Proceedings of the tenth ACM international conference on future energy systems},
  pages={52--57},
  year={2019}
}

@inproceedings{batra2013s,
  title={It's Different: Insights into home energy consumption in India},
  author={Batra, Nipun and Gulati, Manoj and Singh, Amarjeet and Srivastava, Mani B},
  booktitle={Proceedings of the 5th ACM workshop on embedded systems for energy-efficient buildings},
  pages={1--8},
  year={2013}
}

@misc{EIA_RECS2020,
  author       = {{U.S. Energy Information Administration}},
  title        = {Residential Energy Consumption Survey (RECS), 2020},
  howpublished = {\url{https://www.eia.gov/consumption/residential/}},
  year         = {2022},
  note         = {U.S. Department of Energy, Washington, DC},
}

\newcommand{\supplementheading}[2]{%
  \clearpage
  \setcounter{section}{0}%
  \setcounter{figure}{0}%
  \setcounter{table}{0}%
  \renewcommand{\thesection}{#1.\Roman{section}}%
  \renewcommand{\thefigure}{#1\arabic{figure}}%
  \renewcommand{\thetable}{#1\arabic{table}}%
  \renewcommand{\theHsection}{appendix.#1.\arabic{section}}%
  \renewcommand{\theHfigure}{appendix.#1.\arabic{figure}}%
  \renewcommand{\theHtable}{appendix.#1.\arabic{table}}%
  \twocolumn[{\noindent\titlefont\textcolor{ieeedata}{APPENDIX #1: #2}\par\vspace{1.5em}}]%
  \phantomsection
  \label{app:#1}%
  \addcontentsline{toc}{section}{Appendix #1: #2}%
}

\supplementheading{A}{SURVEY METHODOLOGY NOTE}
\newcolumntype{R}{>{\raggedleft\arraybackslash}X} 

\section{OVERVIEW}

This Survey Methodology Note provides a detailed account of the household survey that complements the dataset described in the main article. It documents the sampling frame, fieldwork procedures, and quality-control protocols, and expands on the main paper by including methodological details such as enumerator recruitment and training, questionnaire development and piloting, back-checking and monitoring, and approaches to managing attrition across survey waves. It is organized into sections covering (I) overall approach and objectives, (II) sampling and participant recruitment, (iii) survey instrument development, (iv) fieldwork administration and (v) quality control measures.

\section{OVERALL APPROACH AND OBJECTIVES}

\subsection{Study Design and Implementation}
The three survey waves were conducted as face-to-face interviews using Computer Assisted Personal Interviewing (CAPI). Fieldwork was implemented by Survey Research Lanka (SRL), a vendor selected through a competitive procurement process. SRL handled scripting, translation, pilot testing, training, and dataset delivery, while LIRNEasia developed the questionnaires, provided training, monitored fieldwork, and liaised with LECO to secure approvals and household contact information. In Wave 1, over 4,000 households across LECO's seven branch areas were surveyed, with data later linked to electricity consumption records to form a combined dataset of household demand and behavioral drivers. A temporary pause in Wave 1 fieldwork (Jan--Jul 2024) shortened the spacing between waves, preventing uniform intervals across households.

\subsection{Objectives}
The primary objective of the survey component was to collect longitudinal household-level data on demographics, physical characteristics of the households, appliance ownership, and energy behaviors, such that these could be linked with high-frequency smart meter data. The resulting dataset is meant to facilitate analysis around experimentation with behavioral nudges, tariff reforms, and other demand-side management solutions relevant to energy affordability, efficiency, and policy in Sri Lanka.

\section{SAMPLING AND PARTICIPANT RECRUITMENT}

\subsection{Survey Target Group}

The target group was households served by LECO, located primarily along the south-western coastal belt of Sri Lanka.

\subsection{Survey Respondents}

In a selected household, the household head was interviewed. If the household head was not available, a suitable alternative was sought out (e.g., spouse or other) who would be able to provide sufficient details on the household characteristics, other members, and usage of electricity).

Effort was made to talk to the same respondent (i.e., household member) in all three waves in order to be more consistent. Wherever it was not possible to meet the same respondent for Wave 2 and Wave 3, a suitable alternative respondent was interviewed.

\subsection{Sample Frame \& Procedure}

The LECO customer database served as the sampling frame. The target sample size was 4,000 households, split evenly between smart-metered and non--smart-metered customers. To achieve this, a total of 7,996 households were drawn using stratified random sampling, first by meter type and then by LECO branch area.

In Wave 1, 4,063 households were successfully surveyed, comprising 1,788 smart-metered and 2,275 non--smart-metered households. Due to attrition, the achieved sample declined to 3,501 households in Wave 2 and 3,398 in Wave 3 (see Table~\ref{tab:retention}).The deviation from the planned 2,000--2,000 split reflects the finite pool of eligible smart-metered households: once refusals exhausted the available list, replacements were not possible. At the study's outset, LECO had been asked to retain smart-meter data for this predefined list, making adherence to it essential for maintaining consistent linkage between survey responses and consumption records.\par A further discrepancy arose between the number of surveyed smart-metered households (1,788) and those with retained smart-meter readings (1,438). For 350 households, data retention failures on LECO's systems resulted in missing consumption records, effectively rendering them non-smart-metered for the purposes of the final dataset.

\subsection{Selection of Sample LECO Customer Households}

The selection of the sample of LECO customers was conducted by LIRNEasia. A pre-defined number of households (customers) were selected from each stratum.

To ensure accurate identification of the selected households during the survey, the field team was provided with crucial information ahead of time, including the name of the customer, address, contact number, meter type, etc.

During Wave 1, the respondent was informed that the study consisted of 3 waves spread across around 12 months, and his / her consent for participation in the study and availabiity (at the same household) in the next 12 months was sought prior to commencement of the interviewing process.

\begin{table*}[!t]
\centering
\small
\setlength{\tabcolsep}{4pt}
\caption{Final sample achievement and retention rate across Wave 1, Wave 2, and Wave 3}
\label{tab:retention}

\begin{tabularx}{\textwidth}{lRRR RRR R RRR R}
\toprule
\textbf{Branch} &
\multicolumn{3}{c}{\textbf{Wave 1}} &
\multicolumn{3}{c}{\textbf{Wave 2}} &
\multicolumn{1}{c}{\shortstack{\textbf{Retention}\\\textbf{(W2 vs.\ W1)}}} &
\multicolumn{3}{c}{\textbf{Wave 3}} &
\multicolumn{1}{c}{\shortstack{\textbf{Overall}\\\textbf{Retention}\\\textbf{(W3 vs.\ W1)}}} \\
\cmidrule(lr){2-4}\cmidrule(lr){5-7}\cmidrule(lr){9-11}\\[-0.6ex]
 & \textbf{SM} & \textbf{NSM} & \textbf{Total}
 & \textbf{SM} & \textbf{NSM} & \textbf{Total}
 & & \textbf{SM} & \textbf{NSM} & \textbf{Total} & \\
\midrule
Galle     & 130  & 367  & \textbf{497}  & 114  & 343  & \textbf{457}  & 92.0\% & 108  & 345  & \textbf{453}  & 91.1\% \\
Kalutara  & 85   & 458  & \textbf{543}  & 68   & 427  & \textbf{495}  & 91.2\% & 67   & 419  & \textbf{486}  & 89.5\% \\
Kelaniya  & 78   & 364  & \textbf{442}  & 61   & 293  & \textbf{354}  & 80.1\% & 46   & 277  & \textbf{323}  & 73.1\% \\
Kotte     & 121  & 226  & \textbf{347}  & 89   & 195  & \textbf{284}  & 81.8\% & 78   & 179  & \textbf{257}  & 74.1\% \\
Moratuwa  & 1157 & 309  & \textbf{1466} & 999  & 275  & \textbf{1274} & 86.9\% & 991  & 267  & \textbf{1258} & 85.8\% \\
Negombo   & 46   & 296  & \textbf{342}  & 37   & 277  & \textbf{314}  & 91.8\% & 36   & 277  & \textbf{313}  & 91.5\% \\
Nugegoda  & 171  & 255  & \textbf{426}  & 129  & 194  & \textbf{323}  & 75.8\% & 117  & 191  & \textbf{308}  & 72.3\% \\
\midrule
\textbf{Total} & \textbf{1788} & \textbf{2275} & \textbf{4063} &
\textbf{1497} & \textbf{2004} & \textbf{3501} & \textbf{86.2\%} &
\textbf{1443} & \textbf{1955} & \textbf{3398} & \textbf{83.6\%} \\
\bottomrule
\end{tabularx}

\par\vspace{2pt}
\footnotesize \textit{Note: SM -- Smart metered, NSM -- Non-smart metered}
\end{table*}

\subsection{Household Replacement Procedures}

In the case that a respondent from the household being not available at the first visit, a minimum of three (3) attempts (first visit + two follow-ups) were made (at different times of day, and different days of the week, or based on an appointment) to reach the household. If the household could still not be reached, the household was considered as non-contactable, and was replaced by another household within the same stratum.

\section{SURVEY INSTRUMENT DEVELOPMENT}

\subsection{Development, Translation \& Scripting}

In keeping with the objectives of the study, in order to capture the temporal changes of consumer behavior and consumption patterns in the selected customer households, three separate questionnaires were developed for the three waves. The full questionnaire, provided in Appendix B, was developed to capture the following information areas.

\textbf{Wave 1}: During the initial visit, the household respondents were informed that they were being included in a longitudinal study and that they will be interviewed for another two waves (in total, three waves) of the survey within the next year, and their consent was taken for participation in the survey.

The following information was also collected in the Wave 1 survey:
\begin{itemize}[noitemsep]
    \item List of household members, including their demographics and their staying habits at home
    \item Housing characteristics
    \item A list of all the rooms/areas in the household, characteristics of each room/ area such as the type of ceiling, floor material, number of doors and windows, ventilation, information related to lighting, AC, etc
    \item List of electrical appliances and usage
    \item Other methods of generating electricity (mainly focusing on solar power)
    \item Behaviour on cooking and water heating
\end{itemize}

\textbf{Wave 2 \& 3}: During the follow-up visits, any changes to the information captured in Wave 1 was captured. i.e., the respondent was questioned on whether any of the above-mentioned characteristics have changed since the previous wave. If any characteristics had changed, the enumerator was required to record the information on the existing state of the household. Apart from the changes to information captured in Wave 1, Wave 2, and Wave 3 also captured some additional information.

The following information was captured in wave 2:
\begin{itemize}[noitemsep]
    \item Changes to list of household members (vs. wave 1); i.e., whether the members who were there in wave 1 are still living in the same household, members who have left/passed away, and new additions, demographics of new additions, and staying habits at home for all members currently living in the household
    \item Changes to housing characteristics (vs. wave 1)
    \item Changes to the list of all the rooms/areas in the household, changes to the characteristics of each room/area, such as the type of ceiling, floor material, number of doors and windows, ventilation, information related to lighting, use of fans in the room/area, and AC, etc. (vs. wave 1)
    \item Changes to the list of electrical appliances and usage (vs. wave 1)
    \item Changes to solar power usage (vs. wave 1)
    \item Details of electrical appliances at home, including brand name, when purchased, etc.
    \item Behaviour related to ironing, washing, and drying clothes, etc.
\end{itemize}

The following information was captured in wave 3:

\begin{itemize}[noitemsep]
    \item Changes to list of household members (vs. wave 2); i.e., whether the members who were there in wave 2 are still living in the same household, members who have left/ passed away, and new additions, demographics of new additions, and staying habits at home for all members currently living in the household
    \item Changes to housing characteristics (vs. wave 2)
    \item Changes to the list of all the rooms/ areas in the household, changes to the characteristics of each room/area, such as the type of ceiling, floor material, number of doors and windows, ventilation, information related to lighting, use of fans in the room/area, and AC, etc. (vs. wave 2)
    \item Changes to the list of electrical appliances and usage (vs. wave 2)
    \item Changes to solar power usage (vs. wave 2)
    \item General behaviour related to the use of electrical devices
    \item Awareness of ``Time of use metering''
    \item Electricity billing and payment habits
\end{itemize}

The English language questionnaires for all three waves were developed by LIRNEasia. The initial wave 1 questionnaire was translated into Sinhala and was pilot tested internally by LIRNEasia research team, via face-to-face interviewing of a few electricity consumers. This pilot testing was carried out to assess the questionnaire length, the flow of the questionnaire and the respondents' understanding of the questions, cognitive difficulties, and question sensitivities. Based on the feedback, the wave 1 questionnaire was refined prior to sharing with SRL. Further, based on the same feedback, the wave 2 and wave 3 questionnaires were also fine-tuned and finalised.

The final questionnaires for all three waves were translated into two languages (Sinhala and Tamil) by SRL. Translations were extensively checked by the LIRNEasia research team.

The scripting of the questionnaires were done by SRL. The SRL project teams tested the script for all logical and consistency checks before planning for pilot-test of the finalised questionnaires. The bi-lingual tools and script with login details were also shared with LIRNEasia team for comments and feedback before the pilot test.

\subsection{Pilot Testing the Final Questionnaire}

The pilot-test of the finalised, CAPI scripted questionnaire was conducted in an actual field setting. The pilot-tests were conducted among LECO customer households, which were not selected for the survey sample.

The key purpose of the pilot survey included:
\begin{itemize}[noitemsep]
    \item Estimate realistic average Length of Interview (LOI).
    \item Test skipping and routing.
    \item Test question phrasing
    \item Test translation (language)
    \item Test understanding of the questions, cognitive difficulties, and question sensitivities.
    \item Test tablet functionality
\end{itemize}

The pilot testing exercise was led by the senior SRL professionals and LIRNEasia researchers on the ground.
Based on the observations made during the pilot tests, various improvements to the field process as well as the research tools were made.

\section{FIELDWORK ADMINISTRATION}

\subsection{Interviewer Training}

Prior to commencement of fieldwork for each wave, a one-day training session was held for all interviewers working on the study. During the training session, the questionnaire was discussed in detail, including definitions of household, how to note down timings, explanations of certain terminology used, etc. Further, for some questions (example: such as number of doors, windows, etc.), the rationale behind capturing this data was explained.

Apart from detailed training on the questionnaire, the interviewers were trained on general Dos and Don'ts in the field, following the ethical guidelines, and emphasis was also made on professional conduct when making appointments with customers, especially on how to handle situations with customers in case any issues arise.

Mock interviews were conducted by enumerators before they were sent to the field for data collection.

The interviewers were asked to make appointments before meeting the customers for the interview. They were also requested to carry all authentication documentation with them at all times when conducting fieldwork.

\subsection{Respondent Consent}

Prior to the commencement of an interview, the respondent was informed of the following:
\begin{itemize}[noitemsep]
    \item The objectives of the research
    \item That his/her participation was voluntary
    \item That he/she could choose to end the interview at any point
    \item That they are being included in a longitudinal study and that they will be interviewed for another two waves (in total, three waves) of the survey within the next year (only in wave 1)
\end{itemize}

\subsection{Conducting Fieldwork}

Fieldwork for Wave 1 began in November 2023 but was temporarily suspended at the end of January 2024 due to concerns raised by LECO. Following consultations, fieldwork resumed in July 2024 and was completed by mid-October 2024.

Due to time limitations due to aforementioned delays, wave 2 was also commenced in September 2024, with an overlap to wave 1. Similarly, wave 3 commenced in the end of November 2024 and progressed till early January 2025.

Due to fieldwork being put on-hold and the tight timelines, it was not possible to keep the same time gap between waves for all households. However, at least a 3-4 weeks gap between each wave was maintained for all households.

\subsection{Response Rate}

The SRL team was given a list of 7,996 contacts at the start of fieldwork, with another 2,000 non-smarter meter household contacts added later due to low hit rate.

The overall success rate during wave 1 was 40.6\%, which was slightly higher among the smart-meter segment (44.7\%). Among the non-smart meter segment, the success rate was 37.9\%. Table~\ref{tab:contacts-wave1} shows the breakdown of all the contacts made for wave 1.

\begin{table}[b]
\centering
\caption{Summary of contacts made | Wave 1}
\label{tab:contacts-wave1}
\begin{tabularx}{\linewidth}{>{\raggedright\arraybackslash}X r r r}
\toprule
 & \thead{\textbf{Smart}\\\textbf{meter}}
 & \thead{\textbf{Non-smart}\\\textbf{meter}}
 & \thead{\textbf{Total}} \\
\midrule
\textbf{Success rate} & \textbf{44.7\%} & \textbf{37.9\%} & \textbf{40.6\%} \\
Total contacts given & 3,996 & 6,000 & 9,996 \\
\textbf{Completed successfully} & \textbf{1,788} & \textbf{2,275} & \textbf{4,063} \\
Appointment / Not visited & 114 & 1,729 & 1,843 \\
Non-residential & 347 & 375 & 722 \\
Closed / Under construction & 172 & 274 & 446 \\
Refused & 1,491 & 1,227 & 2,718 \\
Wrong Address & 84 & 120 & 204 \\
\bottomrule
\end{tabularx}
\end{table}

Table~\ref{tab:contacts-wave2} summarises the contacts made for wave 2. Of the total housweholds contacted in wave 1, only 3,501 (86.2\%) were re-contactable for wave 2, which resulted in an attrition of 13.8\%.

\begin{table}[h]
\centering
\caption{Summary of contacts made | Wave 2}
\label{tab:contacts-wave2}
\begin{tabularx}{\linewidth}{>{\raggedright\arraybackslash}X r r r}
\toprule
 & \thead{\textbf{Smart}\\\textbf{meter}}
 & \thead{\textbf{Non-smart}\\\textbf{meter}}
 & \thead{\textbf{Total}} \\
\midrule
\textbf{Total completed in W1} & \textbf{1,788} & \textbf{2,275} & \textbf{4,063} \\
Retention in W2 & 83.7\% & 88.1\% & 86.2\% \\
\textbf{W2 Contact Summary} & & & \\
\textbf{Completed successfully} & \textbf{1,788} & \textbf{2,275} & \textbf{4,063} \\
Appointment / Not Visited & 41 & 39 & 80 \\
Change in resident members & 7 & 9 & 16 \\
Closed / Under Construction & 46 & 37 & 83 \\
Refused & 197 & 186 & 383 \\
Total & 1,788 & 2,275 & 4,063 \\
\bottomrule
\end{tabularx}
\end{table}

Of the households contacted in Wave 2, 97.1\% of the households were successfully re-contacted for Wave 3 (i.e.; 3,398 households). Among smart meter households, the retention was 96.4\% whereas among non-smart meter households, the retention was 97.6\%. Table~\ref{tab:contacts-wave3} summarises the contacts made for wave 3.

At the end of wave 3, a total of 3,398 households were sampled, of which, 1,443 were smart-metered and the remaining 1,955 were non-smart metered.

As explained in the previous section (Response rate), of the 4,063 households contacted in wave 1, a total of 3,398 were retained in wave 3, resulting in an overall retention rate of 83.6\%. The retention was high in areas falling under the LECO branches in Negombo and Galle while Nugegoda branch area had the lowest retention, followed by Kelaniya and Kotte branch areas.

Table~\ref{tab:retention} summarises the final sample achievement across the three waves by each branch area, together with the retention rate.

\begin{table}[t]
\centering
\caption{Summary of contacts made | Wave 3}
\label{tab:contacts-wave3}
\begin{tabularx}{\linewidth}{>{\raggedright\arraybackslash}X r r r}
\toprule
 & \thead{\textbf{Smart}\\\textbf{meter}}
 & \thead{\textbf{Non-smart}\\\textbf{meter}}
 & \thead{\textbf{Total}} \\
\midrule
\textbf{Total completed in W2} & \textbf{1,497} & \textbf{2,004} & \textbf{3,501} \\
Retention in W3 & 96.4\% & 97.6\% & 97.1\% \\
\textbf{W2 Contact Summary} & & & \\
\textbf{Completed successfully} & \textbf{1,443} & \textbf{1,955} & \textbf{3,398} \\
Appointment / Not Visited & 1 & 6 & 7 \\
Change in resident members & - & - & - \\
Closed / Under Construction & 13 & 8 & 21 \\
Refused & 40 & 35 & 75 \\
Total & 1,497 & 2,004 & 3,501 \\
\bottomrule
\end{tabularx}
\end{table}

\section{QUALITY CONTROL (QC) MEASURES}

All research processes and practices followed at SRL are based on ESOMAR research guidelines. Following are some of the QC procedures that were adopted for the survey by SRL and LIRNEasia:
\begin{itemize}[noitemsep]
    \item Accompaniments | Supervisors accompanied the first few interviews of each interviewer.
    \item Back-checks | 18\% of each interviewer's interviews were back-checked by supervisors.
    \item Spot Checks | 10\% of the sample was spot checked by the research team.
    \item Telephonic Back Checks | Once the data was synced to the server, telephonic back checks were done for ~80\% of the interviews.
    \item Voice Checks | Randomly recorded parts of more than 60\% of interviews were listened to, to check whether the interviewer carried out the interview as briefed.
    \item Map Checks | GPS locations of 100\% of the interviews were captured and checked for authenticity of the selected respondent.
    \item For all interviews, interview length was checked to ensure the duration of each interview and that it aligns with expected standards, and to identify any discrepancies.
    \item CAPI script ensured the right filters for the questionnaire.
    \item Once the data was collected, it was checked for any outliers.
    \item Debriefing sessions were done for interviewers who made mistakes during the interview process.
\end{itemize}

\supplementheading{B}{SURVEY QUESTIONNAIRE}
\newcounter{qcounter}
\providecommand{\theHqcounter}{}
\renewcommand{\theHqcounter}{\theHsection.\arabic{qcounter}}
\newcommand{\question}[1]{%
  \refstepcounter{qcounter}%
    \vspace{0.5em}%
  \noindent\textbf{Q\theqcounter. #1}\par
}

\section{Survey Wave 1}

\question{Is this premise used as a household or for business/commercial purposes?}

\question{Are you the household head? By household head, I mean the person who makes the most important financial decisions in the house/the person who contributes the maximum of HH day-to-day expenses.}

\question{Are you able to accurately tell me about the details related to the household?}

\question{Do you consent to participate in this study and to provide your data to LIRNEasia?}

\question{May we audio record some parts of this interview for quality checking purposes?}

\question{Can you please tell me how many electricity meters are in your house?}
\begin{enumerate}[label=(\alph*), nosep, leftmargin=*]
  \item 1 meter
  \item 2 meters
  \item More than 2 meters
\end{enumerate}

\question{Do you or any member of your household own this house, or are you living on rent?}
\begin{enumerate}[label=(\alph*), nosep, leftmargin=*]
  \item Yes, I or a household member owns it.
  \item No, I am living on rent, and I or a household member pays the rent.
  \item No, I am living on rent, and the employer pays the rent.
  \item No, I or any household member does not own or rent this household. We occupy this household without any payment of rent.
\end{enumerate}

\question{Do you occupy any of the following people in your house? By house, we mean the area covered by your electricity meter.}
\begin{enumerate}[label=(\alph*), nosep, leftmargin=*]
  \item Renters/boarders who are living in your annexe or any other attached place, maintaining separate living conditions but sharing the same electricity meter.
  \item Boarders who live in your house using a room/s that are attached to your living conditions.
  \item I don't occupy any of the above.
\end{enumerate}

\question{Which of the following best describes your awareness of the electricity consumption of renters/boarders? Are you aware of details such as the appliances they use and the number of hours they use each appliance, the times they keep the lights and fans switched on, etc.?}
\begin{enumerate}[label=(\alph*), nosep, leftmargin=*]
  \item I know all the details about the electricity consumption of the renters/ boarders; i.e., the appliances they use and the number of hours they use each appliance, the times they keep the lights and fans switched on, etc.
  \item I know some details about the electricity consumption of the renters/ boarders; i.e., the appliances they use and the number of hours they use each appliance, the times they keep the lights and fans switched on, etc.
  \item I do not know any details about the electricity consumption of the renters/ boarders; i.e, the appliances they use and the number of hours they use each appliance, the times they keep the lights and fans switched on, etc.
\end{enumerate}

\question{Which of the following time periods best describes when your house was built?}
\begin{enumerate}[label=(\alph*), nosep, leftmargin=*]
    \item Before 1980
    \item 1980-1989
    \item 1990-1999
    \item 2000-2009
    \item 2010-2019
    \item In 2020 or After 2020
    \item Don't know
\end{enumerate}

\question{Select the best option that describes your house.}
\begin{enumerate}[label=(\alph*), nosep, leftmargin=*]
    \item Single House - Single Floor
    \item Single House - Double Floor
    \item Single House - More than 2 floors
    \item Attached house / Annex
    \item Flat
    \item Condominium / Luxury apartments
    \item Twin houses
    \item Line room / Row house
    \item Slum / Shanty
    \item Other
\end{enumerate}

\question{What floor is your house located on? (If the house occupies multiple floors, the answer should be the lowest floor}

\question{How many storeys does your house have?}

\question{Can you please tell me what the floor area of this household is? i.e., how many square feet does this household cover? Please remember, when we say household, what we mean is the parts of the household that are covered by the electricity meter.}

\question{Please tell me the number of members living in this household. By household, I mean the area covered by the electricity meter. If there are members who are temporarily living abroad or living in some other part of Sri Lanka, please include them as well. Also, if there are any boarders or those who are living on rent, or if you have any maids/housekeepers staying in this household, please include them as well.}

\question{Name of the household member}

\question{Relationship to the household head}
\begin{enumerate}[label=(\alph*), nosep, leftmargin=*]
    \item Head of the household
    \item Wife/Husband
    \item Son/daughter
    \item Son-in-law/Daughter-in-law
    \item Parents of the head of the household/spouse
    \item Grandson/Granddaughter
    \item Other relative
    \item Domestic servant/driver/watcher
    \item Boarder
    \item Other
\end{enumerate}

\question{Gender}
\begin{enumerate}[label=(\alph*), nosep, leftmargin=*]
    \item Male
    \item Female
    \item Other
\end{enumerate}

\question{Year of birth}
\question{Age}

\question{Ethnicity}
\begin{enumerate}[label=(\alph*), nosep, leftmargin=*]
    \item Sinhala
    \item Sri Lankan Tamil
    \item Indian Tamil
    \item Sri Lankan Moor/Muslim
    \item Burgher
    \item Malay
    \item Other
\end{enumerate}

\question{Religion}
\begin{enumerate}[label=(\alph*), nosep, leftmargin=*]
    \item Buddhism
    \item Hinduism
    \item Islam
    \item Roman Catholicism
    \item Other Christian denominations
    \item No religion
    \item Other
\end{enumerate}

\question{Marital status}
\begin{enumerate}[label=(\alph*), nosep, leftmargin=*]
    \item Never married
    \item Currently married (registered)
    \item Currently married (customary)
    \item Widowed
    \item Divorced
    \item Legally separated
    \item Separated (not legally)
    \item Not married but lives as a family
    \item Other
\end{enumerate}

\question{Current attendance at school or any other educational institution. i.e., is he/she currently doing any type of studying? This could be a degree, an MBA, etc., or following a course like CIMA, CIM, or a vocational training course?}
\begin{enumerate}[label=(\alph*), nosep, leftmargin=*]
    \item Preschool
    \item School
    \item University
    \item Other educational institution
    \item Vocational/Technical Institution
    \item Pending results G.C.E. (O.L / A.L)
    \item Still a toddler
    \item Does not attend
\end{enumerate}

\question{Highest level of education}
\begin{enumerate}[label=(\alph*), nosep, leftmargin=*]
    \item Studying/Studied Grade 1
    \item Passed Grade 1
    \item Passed Grade 2
    \item Passed Grade 3
    \item Passed Grade 4
    \item Passed Grade 5
    \item Passed Grade 6
    \item Passed Grade 7
    \item Passed Grade 8
    \item Passed Grade 9
    \item Passed Grade 10
    \item Passed G.C.E. (O/L)
    \item Passed Grade 12
    \item Passed G.C.E. (A/L) or equivalent
    \item Passed GAQ (General Arts Qualifying (External) Examination) / GSQ (General Science Qualifying Examination)
    \item Passed Degree / Diploma
    \item Passed Post Graduate Degree / Diploma
    \item PhD
    \item Special Education learning / learnt
    \item No Schooling
\end{enumerate}

\question{Main activity usually engaged in}
\begin{enumerate}[label=(\alph*), nosep, leftmargin=*]
    \item Engaged in economic activity / currently employed/engaged in own business
    \item Retired and obtaining government/semi-government pension payment and is currently engaged in economic activity (employed elsewhere other than the place where he/she is receiving the pension from / engaged in own business)
    \item Seeking and available to work
    \item Retired - Obtaining government/semi-government pension payment and currently not engaged in economic activity (not employed elsewhere or not engaged in any own business)
    \item Retired from the private/semi-government sector and does not receive any pension payment.
    \item Received other pension payments.
    \item Household activities
    \item Student
    \item Too old / Disabled/unable to work.
    \item Other
\end{enumerate}

\question{Main occupation}
\begin{enumerate}[label=(\alph*), nosep, leftmargin=*]
    \item Legislator, senior official, and manager
    \item Professional
    \item Technician and associate professional
    \item Clerk
    \item Service worker, and shop and market sales worker
    \item Skilled agricultural and fishery worker
    \item Craft and related worker
    \item Plant and machine operator and assembler
    \item Elementary occupation
    \item Related to forces
    \item No occupation
\end{enumerate}

\question{Daily wage owner or not}

\question{Employment status of the main occupation}
\begin{enumerate}[label=(\alph*), nosep, leftmargin=*]
    \item Government employee
    \item Semi-government employee
    \item Private sector employee
    \item Employer
    \item Own account worker
    \item Contributing family worker
\end{enumerate}

\question{The electricity consumption can vary based on the time that people stay at home. Therefore, we would like to know the number of hours spent at home during the last week for each member.}

\question{During the last week, did this member go out of the home for employment purposes?}
\begin{enumerate}[label=(\alph*), nosep, leftmargin=*]
    \item Yes, went daily during working days.
    \item Yes, went on most of the days.
    \item No, worked from home.
\end{enumerate}

\question{You mentioned that renters or boarders are living in this household who share the same electricity meter. Which of the following best describes how you charge them for electricity?}
\begin{enumerate}[label=(\alph*), nosep, leftmargin=*]
    \item You charge a fixed amount every month for electricity.
    \item You charge an amount for electricity depending on the variance of the bill.
    \item You don't charge a specific amount for electricity, but charge a fixed amount for all the utilities, such as electricity, water, etc.
    \item You don't charge a specific amount for electricity, but charge a varied amount for all the utilities, such as electricity, water, etc. The amount charged varies based on the utility bills.
    \item You don't charge them for electricity consumption.
\end{enumerate}

\question{You mentioned that you are living on rent. Which of the following best describes the payment for your electricity consumption?}
\begin{enumerate}[label=(\alph*), nosep, leftmargin=*]
    \item You pay the full amount of the electricity bill.
    \item You pay a fixed amount to the owner every month for electricity.
    \item You pay a varied amount to the owner every month for electricity. The amount paid varies depending on the variance of the bill.
    \item You don't pay a specific amount for electricity, but pay a fixed amount for all the utilities, such as electricity, water, etc.
    \item You don't pay a specific amount for electricity, but pay a varied amount for all the utilities, such as electricity, water, etc. The amount paid varies depending on the variance of the utility bills.
    \item You don't pay the owner for electricity consumption.
\end{enumerate}

\question{Is there any business activity that is being carried out in any part of this household, such as a shop, a communication, your own office, etc., for which electricity is used from the same meter?}

\question{What type of business activity is being carried out in this household?}
\begin{enumerate}[label=(\alph*), nosep, leftmargin=*]
    \item A shop
    \item A communication
    \item Other
\end{enumerate}

\question{Do you have any of these appliances in your household in working condition, which you have used at least once? If you have boarders or those who are living on rent in your household, please include the electrical appliances used by them as well.}

\question{Number of appliances of the type}

\question{Duration of usage in hours during the last week}

\question{Can you please tell me which of these best describes by whom or how your house was designed?}
\begin{enumerate}[label=(\alph*), nosep, leftmargin=*]
    \item The house is designed by a certified architect.
    \item The house plan was not done by an architect but checked by a certified architect or engineer.
    \item The house plan is not done by an architect, nor checked by a certified architect/engineer; the house is designed to barely pass the legal requirements of the local authorities.
    \item The house plan is not done by an architect, nor checked by a certified architect or engineer. The house was not designed keeping in mind the legal requirements of the local authorities; it was designed only to suit our needs.
    \item This is a house provided by the government.
    \item I am not aware of that.
\end{enumerate}

\question{Do you have a COC (Certificate of Compliance) for your home? (A compliance certificate is a certificate given by the municipal council or the local council after coming to the house and inspecting it.)}

\question{What is the main material used to build the outside walls of your house?}
\begin{enumerate}[label=(\alph*), nosep, leftmargin=*]
    \item Brick
    \item Cement Block
    \item Stones/Cube stones
    \item Cabook
    \item Pressed soil blocks
    \item Cadjan / Palmyra
    \item Wood / Takaran / Asbestos
    \item Metal Sheet
    \item Mud
    \item Other
    \item I am not aware of that.
\end{enumerate}

\question{What is the main material used for most of the roof of your house? Please remember, when we say household, what we mean is the parts of the household that are covered by the electricity meter.}
\begin{enumerate}[label=(\alph*), nosep, leftmargin=*]
    \item Tile
    \item Asbestos
    \item Concrete
    \item Metal Sheet
    \item Takaran
    \item Cadjun/Palmyrah/Straw
    \item Tent
    \item Plastic sheets
    \item Other
\end{enumerate}

\question{We would like to know more about the different types of rooms and different sections in your house. When we say house, we mean the area covered by the electricity meter. How many of the following types of rooms are there in your house? If there are rooms or parts of the household that are not directly connected to the house but are still covered by the same electricity meter, such as a separate kitchen or a toilet, please include these as well.}
\begin{enumerate}[label=(\alph*), nosep, leftmargin=*]
    \item Living room
    \item Bedrooms
    \item Kitchen and/ or pantry/dining room
    \item Bathroom and/or toilets
    \item Storage room
    \item Other rooms/parts in the house, including passages, verandas, balconies, staircase, etc.
    \item Garage
    \item Security Room
    \item Garden (including the wall around the garden, if any)
    \item Other rooms outside the house
\end{enumerate}

\question{For what purposes do you mainly use this section of the household?}
\begin{enumerate}[label=(\alph*), nosep, leftmargin=*]
    \item Living room
    \item Bedrooms
    \item Kitchen and/ or pantry
    \item Bathroom and/or toilets
    \item Storage room
    \item Gaming room
    \item Servant's Room
    \item Visitors' room
    \item Passage
    \item Veranda
    \item Balcony
    \item Staircase
    \item Garage
    \item Security Room
    \item Garden
    \item Other rooms outside the house
    \item Study room /Office room
    \item Other
\end{enumerate}

\question{In which storey is this section of the household located? (Define ground floor by zero)}

\question{What is the main material used for the roof of this section of the household?}
\begin{enumerate}[label=(\alph*), nosep, leftmargin=*]
    \item Tile
    \item Asbestos
    \item Concrete
    \item Metal Sheet
    \item Takaran
    \item Cadjun/Palmyra/Straw
    \item Tent
    \item Plastic sheets
    \item Other
    \item Garden -- Not relevant
\end{enumerate}

\question{What type of ceiling does this section of the household have?}
\begin{enumerate}[label=(\alph*), nosep, leftmargin=*]
    \item No ceiling, just the roof above
    \item A conventional ceiling
    \item A beamed ceiling
    \item A hanging ceiling
    \item A polythene cover as a ceiling
    \item No ceiling, the concrete slab
    \item Wooden ceiling
    \item Other
    \item Garden -- Not relevant
\end{enumerate}

\question{What is the main material used for the floor of this section of the household?}
\begin{enumerate}[label=(\alph*), nosep, leftmargin=*]
    \item Cement
    \item Teraso
    \item Tile
    \item Granite
    \item Wood (finished)
    \item Mud
    \item Wood
    \item Sand
    \item Concrete
    \item Other
    \item Garden -- Not relevant
\end{enumerate}

\question{How many of the doors of this section of the household open to the garden/external environment?}

\question{How many windows in this section of the household can be opened and closed to the outside?}

\question{What is the main material used for the window panes?}
\begin{enumerate}[label=(\alph*), nosep, leftmargin=*]
    \item None, it's open
    \item Wood
    \item Glass
    \item Net
    \item Other
    \item Garden -- Not relevant
\end{enumerate}

\question{Do you have curtains or blinds for the windows?}

\question{Does this section of the household have natural ventilation other than windows, to allow fresh air circulation? This could be a ventilation hole in the wall or an area with a grill on top of the windows, which is not covered by the window pane, etc.}

\question{How many light bulbs do you have in this section of the household?}

\question{How many fans do you have in this section of the household? Please include all types of fans such as ceiling fans, pedestal fans, table fans, exhaust fans, tower fans, etc.}

\question{How many ACs do you have in this section of the household?}

\question{What is the type of the light Bulb out of the following?}
\begin{enumerate}[label=(\alph*), nosep, leftmargin=*]
    \item Incandescent
    \item CFL
    \item LED
    \item Halogen
    \item Tube Light (conventional)
    \item Tube Light (LED)
    \item Flashlight
    \item Flood light
    \item Other
\end{enumerate}

\question{What is the wattage (W) of the light?}

\question{For how many hours was this light kept switched on last week during the daytime. (8 am --6 pm)?}

\question{For how many hours was this light kept switched on last week during nighttime. (6 pm -- 8 am)?}

\question{What is the type of the fan?}
\begin{enumerate}[label=(\alph*), nosep, leftmargin=*]
    \item Ceiling fan
    \item Wall fan
    \item Pedestal fan/ Stand fan
    \item Table fan
    \item Exhaust fan
    \item Tower fan
    \item Other
\end{enumerate}

\question{For how many hours was this fan kept switched on last week during the daytime. (8 am -- 6 pm)?}

\question{For how many hours was this fan kept switched on last week during nighttime. (6 pm -- 8 am)?}

\question{What is the type of air conditioning used in the room?}
\begin{enumerate}[label=(\alph*), nosep, leftmargin=*]
    \item Central AC (Only to your Household)
    \item Central AC (For the whole building. Relevant if B.5 = 5 or 6)
    \item Individual AC with two components
    \item Individual AC with one component
    \item Air Cooler
    \item Other
\end{enumerate}

\question{Is the AC an inverter AC?}

\question{Which of the following is true regarding air conditioners?}
\begin{enumerate}[label=(\alph*), nosep, leftmargin=*]
    \item The room can be fully closed and sealed, and there are no outside openings. When the AC is turned on, the cool air does not go out of the room.
    \item The room is fully closed. However, it is not fully sealed. Therefore, when the AC is on, the cool air may leak through the spaces that are not sealed, such as the space in-between the door and the door frame
    \item The room is not fully closed. There are spaces where the cool air can leak out
\end{enumerate}

\question{What is the wattage of the AC?}

\question{What is the BTU of the AC?}

\question{For how many hours was this AC kept switched on, last week, during daytime? (8 am --6 pm)?}

\question{For how many hours was this AC kept switched on last week, during nighttime. (66 pm-- 8 am)?}

\question{Are there any other parts of the household which are covered by the same electricity meter that were not covered above?}

\question{Can you please tell me which parts of the house you have missed mentioning earlier? Please mention only the parts you have missed, and to which electricity is supplied from the same electricity meter?}

\question{Are there any constructions or renovations happening in this household currently?}

\question{Does your household have a backup generator to generate electricity if needed?}

\question{Apart from the electricity that you get from the national grid and the back-up generator, do you use any other methods for generating electricity at your household? If so, what methods?}
\begin{enumerate}[label=(\alph*), nosep, leftmargin=*]
    \item Solar Energy
    \item Bioenergy
    \item Mini Hydropower
    \item Wind Power
    \item Other
\end{enumerate}

\question{Is your solar system on-grid or off-grid?}

\question{Do you have an inverter for the solar energy generation system? An inverter is a device that converts direct current (DC) electricity, which is what a solar panel generates, to alternating current (AC) electricity, which the electrical grid uses.}

\question{Which of the following purposes do you use solar energy for in your house?}
\begin{enumerate}[label=(\alph*), nosep, leftmargin=*]
    \item Water heating
    \item Cooking
    \item Outdoor lighting
    \item Car charging
    \item Agriculture equipment and systems (irrigation systems, etc.)
    \item All the above
    \item Other
\end{enumerate}

\question{Are you aware of the number of units generated by the solar energy system in the last month?}

\question{How many units did the solar system generate last month?}

\question{When did you start using solar for electricity generation? Was it before November 2022 or after November 2022?}

\question{Do you have a system to backup/store electricity to be used in case of a power cut? This could be a battery with an inverter where you can use the normal lights and appliances of the full household, or a part of the household from the electricity stored in the battery? Please note that what we are referring to here is not the emergency lights.}

\question{How does your household receive water?}
\begin{enumerate}[label=(\alph*), nosep, leftmargin=*]
    \item Protected well.
    \item Unprotected well.
    \item Tube well
    \item Tap Water (National Water Supply and Drainage Board)
    \item Tap Water (Community-based water supply and management organization)
    \item Tap Water (Local Government Organizations)
    \item Tap Water (Private Water Projects)
    \item River/ Tank/ Streams
    \item Rainwater
    \item Bottled water
    \item Bowser
    \item Other
\end{enumerate}

\question{In the case of using hot water for bathing or having a body wash, what would you or any of the household members mostly do?}
\begin{enumerate}[label=(\alph*), nosep, leftmargin=*]
    \item We have an in-built water heating system powered solely by solar energy.
    \item We have an in-built water heating system powered solely by the grid supply.
    \item We have an in-built water heating system powered by both solar and the grid supply.
    \item We have a water heating system powered by a different source.
    \item We don't have an inbuilt system for water heating; we use heated water through an electric kettle/heater.
    \item We don't have an inbuilt system for water heating; we use water heated by means such as gas, firewood, etc. (other than an electric kettle, electric water heater, or water heated using any other electrical appliance).
    \item None, we do not use hot water for bathing or body wash purposes
\end{enumerate}

\question{Does your water heating equipment serve other housing units?}

\question{Did any of your household members use hot water for bathing in the last week?}

\question{Do you normally boil water before drinking?}

\question{What source of energy is used to boil water for drinking?}
\begin{enumerate}[label=(\alph*), nosep, leftmargin=*]
    \item Gas
    \item Electricity (directly from the national grid)
    \item Electricity (generated from solar energy system)
    \item Firewood
    \item Kerosene
    \item Sawdust/ Paddy husk.
    \item Biogas
    \item Coconut shells/charcoal
    \item Other
\end{enumerate}

\question{How many times did you cook meals at home last week?}
\begin{enumerate}[label=(\alph*), nosep, leftmargin=*]
    \item Did not cook at home last week
    \item 1-7 times
    \item 8 -- 14 times
    \item 15 -- 21 times
    \item More than 21 times
\end{enumerate}

\question{What sources of energy does your house use for cooking?}
\begin{enumerate}[label=(\alph*), nosep, leftmargin=*]
    \item Gas
    \item Electricity (directly from the national grid)
    \item Electricity (generated from solar energy system)
    \item Firewood
    \item Kerosene
    \item Sawdust/ Paddy husk.
    \item Biogas
    \item Coconut shells/charcoal
    \item Other
\end{enumerate}

\question{Could you also tell me the highest level of education attained by the chief wage earner in your household? By chief wage earner, I mean the person who contributes most to the household expenses.}
\begin{enumerate}[label=(\alph*), nosep, leftmargin=*]
    \item Illiterate
    \item Primary Education
    \item Schooling up to Grade 6 - 9
    \item O/L or A/L pending / Passed
    \item Diploma with O/L or A/L (Non graduate)
    \item Other professional certificates with O/L or A/L / Part qualification (Non graduate)
    \item Graduate /Post-Grads/ Degree level professional qualification
\end{enumerate}

\question{What is the Occupation of the Chief Wage Earner in your house?}
\begin{enumerate}[label=(\alph*), nosep, leftmargin=*]
    \item Unskilled Worker
    \item Skilled Worker
    \item Clerk / Salesman grades
    \item Supervisor grades
    \item Junior executive / Executive
    \item Middle and Senior executive
    \item Manager / Professional
    \item Small Businessman / Self-employed (Non-professional)
    \item Boutique owner
    \item Self-employed (Professional) - No employees
    \item 1-9 Employed
    \item 10+ Employed
    \item Agricultural labourer / Worker
    \item Tenant cultivator
    \item Farmer owning - Less than \textonehalf{} Acre
    \item Farmer owning - \textonehalf{} - 1 / Acre
    \item Farmer owning - 1 -- 2 / Acre
    \item Farmer owning - 2 -- 5 / Acre
    \item Farmer owning - Over 5 acres / Landed proprietor
\end{enumerate}

\question{What is the total expenditure of your household in the last month? Please consider the spending by all members in your household. Please include all types of expenses, such as food items that you or your household members bought such as  rice, coconut, chillies, sugar, prepared food that you'll bought from outside, you and your household members' spending on personal care products such as soaps, shampoo, toothpaste etc., expenses on laundry products such as washing powder, washing soap, expenses on travelling, on education such as school fees or tuition fees etc., your spending on clothes etc.}

\question{As you know, we are doing this study regarding the electricity consumption. Therefore, can you please tell me what type of electricity meter you have at home? Is it a smart meter or a non-smart meter?}

\section{Survey Wave 2}
\setcounter{qcounter}{0}

\question{Is this premise used as a household or for business/commercial purposes?}

\question{Are you the same respondent as in wave 1, or is it a different member of the household?}
\begin{enumerate}[label=(\alph*), nosep, leftmargin=*]
    \item Same respondent as in wave 1
    \item A different respondent compared to wave 1
\end{enumerate}

\question{Name of respondent}

\question{Are you able to accurately tell me about the details related to the household?}

\question{Do you consent to participate in this study and to provide your data to LIRNEasia?}

\question{May we audio record some parts of this interview for quality checking purposes?}

\question{Have you obtained any new electricity meter/s since we did the last survey?}

\question{Please tell me how many electricity meters are there in your house now?}

\question{Do you or any member of your household own this house, or are you living on rent?}
\begin{enumerate}[label=(\alph*), nosep, leftmargin=*]
    \item Yes, I or a household member owns it.
    \item No, I am living on rent, and I or a household member pays the rent.
    \item No, I am living on rent, and the employer pays the rent.
    \item No, I or any household member does not own or rent this household. We occupy this household without any payment of rent.
\end{enumerate}

\question{Do you occupy any of the following people in your house? By house, we mean the area covered by your electricity meter.}
\begin{enumerate}[label=(\alph*), nosep, leftmargin=*]
    \item Renters/boarders who are living in your annexe or any other attached place, maintaining separate living conditions but sharing the same electricity meter.
    \item Boarders who live in your house using a room/s that are attached to your living conditions.
    \item I don't occupy any of the above.
\end{enumerate}

\question{You said that renters/boarders are living in this house. Can you please tell me whether they were living in this house when we did the 1st interview, or did they start living here afterwards?}
\begin{enumerate}[label=(\alph*), nosep, leftmargin=*]
    \item They were living in this household when the 1st interview was done
    \item They started living in this household after the 1st interview was done
    \item Some were there when the first interview was done, but some are new
\end{enumerate}

\question{Since when did the new renters/boarder(s) start living in this household?}
\begin{enumerate}[label=(\alph*), nosep, leftmargin=*]
    \item December 2023
    \item January 2024
    \item February 2024
    \item March 2024	5. April 2024
    \item May 2024
    \item June 2024
    \item July 2024
\end{enumerate}

\question{Which of the following best describes your awareness of the electricity consumption of renters/boarders?  Are you aware of details such as the appliances they use and the number of hours they use each appliance, the times they keep the lights and fans switched on, etc.?}
\begin{enumerate}[label=(\alph*), nosep, leftmargin=*]
    \item I know all the details about the electricity consumption of the renters/ boarders; i.e., the appliances they use and the number of hours they use each appliance, the times they keep the lights and fans switched on, etc.
    \item I know some details about the electricity consumption of the renters/ boarders; i.e., the appliances they use and the number of hours they use each appliance, the times they keep the lights and fans switched on, etc.
    \item I do not know any details about the electricity consumption of the renters/ boarders; i.e, the appliances they use and the number of hours they use each appliance, the times they keep the lights and fans switched on, etc.
\end{enumerate}

\question{Have there been any structural changes to your household after we did the 1st interview with you? By household, we mean the area covered by your electricity meter. i.e., have there been any additions to the house, such as adding a new part to the household like a room, etc., or completing a section which was being built at the time we did the first interview, or added a new floor, etc.?  or, have there been any parts of the household which were dismantled?}
\begin{enumerate}[label=(\alph*), nosep, leftmargin=*]
    \item Yes, new parts were added only
    \item Yes, some parts were dismantled only
    \item Yes, some new parts were added to the household, and some parts were dismantled
    \item No structural changes were done to the household
\end{enumerate}

\question{Were there any renovations done after we did the 1st interview with you?}

\question{Which of these best describes your house now?}
\begin{enumerate}[label=(\alph*), nosep, leftmargin=*]
    \item Single House - Single Floor
    \item Single House - Double Floor
    \item Single House -- More than 2 floors
    \item Attached house / Annex.
    \item Flat
    \item Condominium/ Luxury apartments
    \item Twin houses
    \item Line room/row house
    \item Slum / Shanty
    \item Other
\end{enumerate}

\question{How many storeys does your house have?}

\question{Can you please tell me what the floor area of this household is now? i.e., how many square feet does this household cover? Please remember, when we say household, what we mean is the parts of the household that are covered by the electricity meter.}

\question{You mentioned that you or a household member owns this house. Can you please tell me whether you've built this house or whether you've purchased it?}
\begin{enumerate}[label=(\alph*), nosep, leftmargin=*]
    \item Built the house
    \item Purchased the house
\end{enumerate}

\question{When you built this house, did you consider any of the following?}
\begin{enumerate}[label=(\alph*), nosep, leftmargin=*]
    \item Natural lighting
    \item Good ventilation
    \item Use of insulation material
\end{enumerate}

\question{Can you please tell me whether there are any changes in the composition of members living in the household? i.e., whether these members are still living in this household or whether they have moved out and hence, are not staying here anymore?}

\question{Apart from the above-mentioned members, are there any others who are currently living in the household? It could be someone who has come back from abroad, or any new additions? It could also be a boarder or a domestic help who has started living in the household after wave 1? If so, how many such members are there?}

\question{Name of the household member}

\question{Relationship to the household head}
\begin{enumerate}[label=(\alph*), nosep, leftmargin=*]
    \item Head of the household
    \item Wife/Husband
    \item Son/daughter
    \item Son-in-law/Daughter in law
    \item Parents of the head of the Household/ spouse
    \item Grandson/ Granddaughter
    \item Other relative
    \item Domestic servant/driver/watcher
    \item Boarder
    \item Other
\end{enumerate}

\question{Gender}
\begin{enumerate}[label=(\alph*), nosep, leftmargin=*]
    \item Male
    \item Female
    \item Other
\end{enumerate}

\question{Year of birth}

\question{Age}

\question{Ethnicity}
\begin{enumerate}[label=(\alph*), nosep, leftmargin=*]
    \item Sinhala
    \item Sri Lankan Tamil
    \item Indian Tamil
    \item Sri Lankan Moor/Muslim
    \item Burgher
    \item Malay
    \item Other
\end{enumerate}

\question{Religion}
\begin{enumerate}[label=(\alph*), nosep, leftmargin=*]
    \item Buddhism
    \item Hinduism
    \item Islam
    \item Roman Catholicism
    \item Other Christian denominations
    \item No religion
    \item Other
\end{enumerate}

\question{Marital status}
\begin{enumerate}[label=(\alph*), nosep, leftmargin=*]
    \item Never married
    \item Currently married (registered)
    \item Currently married (customary)
    \item Widowed
    \item Divorced
    \item Legally separated
    \item Separated (not legally)
    \item Not married but lives as a Family
    \item Other
\end{enumerate}

\question{Current attendance at school or any other educational institution. i.e., is he/she currently doing any type of studying? This could be a degree, an MBA, etc., or following a course like CIMA, CIM, or a vocational training course?}
\begin{enumerate}[label=(\alph*), nosep, leftmargin=*]
    \item Preschool
    \item School
    \item University
    \item Other educational institution
    \item Vocational/Technical Institution
    \item Pending results G.C.E. (O.L / A.L)
    \item Still a toddler
    \item Does not attend
\end{enumerate}

\question{Highest level of education}
\begin{enumerate}[label=(\alph*), nosep, leftmargin=*]
    \item Studying/Studied Grade 1
    \item Passed Grade 1
    \item Passed Grade 2
    \item Passed Grade 3
    \item Passed Grade 4
    \item Passed Grade 5
    \item Passed Grade 6
    \item Passed Grade 7
    \item Passed Grade 8
    \item Passed Grade 9
    \item Passed Grade 10
    \item Passed G.C.E.(O/L)
    \item Passed Grade 12
    \item Passed G.C.E.(A/L) or equivalent
    \item Passed GAQ (General Arts Qualifying (External) Examination)/GSQ (General Science Qualifying examination)
    \item Passed Degree / Diploma
    \item Passed Post Graduate Degree / Diploma
    \item PhD
    \item Special Education learning / learnt.
    \item No Schooling
\end{enumerate}

\question{Main activity usually engaged in}
\begin{enumerate}[label=(\alph*), nosep, leftmargin=*]
    \item Engaged in economic activity/ currently employed/ engaged in own business
    \item Retired and obtaining government/semi-government pension payment and is currently engaged in economic activity (employed elsewhere other than the place where he/she is receiving the pension from / engaged in own business)
    \item Seeking work and available to work
    \item Retired - Obtaining government/semi-government pension payment and currently not engaged in economic activity (not employed elsewhere or not engaged in any own business)
    \item Retired from the private/semi-government sector and does not receive any pension payment.
    \item Received other pension payments.
    \item Household activities
    \item Student
    \item Too old / Disabled/ unable to work.
    \item Other
\end{enumerate}

\question{Main occupation}
\begin{enumerate}[label=(\alph*), nosep, leftmargin=*]
    \item Legislator, senior official, and manager
    \item Professional
    \item Technician and associate professional
    \item Clerk
    \item Service worker, and shop and market sales worker
    \item Skilled agricultural and fishery worker
    \item Craft and related worker
    \item Plant and machine operator and assembler
    \item Elementary occupation
    \item Related to forces.
    \item No occupation
\end{enumerate}

\question{Daily wage owner or not}

\question{Employment status of the main occupation}
\begin{enumerate}[label=(\alph*), nosep, leftmargin=*]
    \item Government employee
    \item Semi-government employee
    \item Private sector employee
    \item Employer
    \item Own account worker
    \item Contributing family worker
\end{enumerate}

\question{The electricity consumption can vary based on the time that people stay at home. Therefore, we would like to know the number of hours spent at home during the last week for each member.}

\question{Was there any change in the main activity usually engaged in? It could be a change in jobs, a promotion, etc., or you started working full-time or part-time, or you started studying full-time, stopped working/retired, etc.}

\question{You mentioned that renters or boarders are living in this household who share the same electricity meter. Which of the following best describes how you charge them for electricity?}
\begin{enumerate}[label=(\alph*), nosep, leftmargin=*]
    \item You charge a fixed amount every month for electricity.
    \item You charge an amount for electricity depending on the variance of the bill.
    \item You don't charge a specific amount for electricity, but charge a fixed amount for all the utilities, such as electricity, water, etc.
    \item You don't charge a specific amount for electricity, but charge a varied amount for all the utilities, such as electricity, water, etc. The amount charged varied based on the utility bills
    \item You don't charge them for electricity consumption
\end{enumerate}

\question{You mentioned that you are living on rent. Which of the following best describes the payment for your electricity consumption?}
\begin{enumerate}[label=(\alph*), nosep, leftmargin=*]
    \item You pay the full amount of the electricity bill.
    \item You pay a fixed amount to the owner every month for electricity.
    \item You pay a varied amount to the owner every month for electricity. The amount paid varies depending on the variance of the bill.
    \item You don't pay a specific amount for electricity, but pay a fixed amount for all the utilities, such as electricity, water, etc.
    \item You don't pay a specific amount for electricity, but pay a varied amount for all the utilities, such as electricity, water, etc. The amount paid varies depending on the variance of the utility bills.
    \item You don't pay the owner for electricity consumption
\end{enumerate}

\question{Is there any business activity that is being carried out in any part of this household, such as a shop, a communication, your own office, etc., for which electricity is used from the same meter?}

\question{What type of business activity is being carried out in this household?}
\begin{enumerate}[label=(\alph*), nosep, leftmargin=*]
    \item A shop
    \item A communication
    \item Other (specify)
\end{enumerate}

\question{You mentioned you had (MENTION THE NUMBER OF APPLIANCE) (MENTION THE CORRESPONDING TYPE OF APPLIANCE).}

\question{Do you still have the same number of (MENTION APPLIANCE TYPE) in working condition?}

\question{No of appliances of the type?}

\question{Duration of usage in hours during the last week}

\question{Apart from the above-mentioned appliances, do you have any of these appliances in your household in working condition?}

\question{Please tell me whether there are any changes to the number of rooms/ parts?}

\question{Are there any changes to the number of (room type)?}
\begin{enumerate}[label=(\alph*), nosep, leftmargin=*]
    \item Living room
    \item Bedrooms
    \item Kitchen and/ or pantry/dining room
    \item Bathroom and/or toilets
    \item Storage room
    \item Other rooms/parts in the house, including passages, verandas, balconies, staircase, etc.
    \item Garage
    \item Security Room
    \item Garden (including the wall around the garden, if any)
    \item Other rooms outside the house
\end{enumerate}

\question{New number of rooms}

\question{For what purposes do you mainly use this section of the household?}
\begin{enumerate}[label=(\alph*), nosep, leftmargin=*]
    \item Living room
    \item Bedrooms
    \item Kitchen and/ or pantry
    \item Bathroom and/or toilets
    \item Storage room
    \item Gaming room
    \item Servant's Room
    \item Visitors' room
    \item Passage
    \item Veranda
    \item Balcony
    \item Staircase
    \item Garage
    \item Security Room
    \item Garden
    \item Other rooms outside the house
    \item Study room /Office room
    \item Other
\end{enumerate}

\question{Was this section a part of the household when we did the first interview in wave 1?}
\begin{enumerate}[label=(\alph*), nosep, leftmargin=*]
    \item Yes, and no changes were made to this section of the household
    \item Yes, but there were some changes made to this section of the household
    \item No, this section was built after the 1st interview
\end{enumerate}

\question{In which storey is this section of the household located?}

\question{What is the main material used for the roof of this section of the household?}
\begin{enumerate}[label=(\alph*), nosep, leftmargin=*]
    \item Tile
    \item Asbestos
    \item Concrete
    \item Metal Sheet
    \item Takaran
    \item Cadjun/Palmyra/Straw
    \item Tent
    \item Plastic sheets
    \item Other
    \item Garden -- Not relevant
\end{enumerate}

\question{What type of ceiling does this section of the household have?}
\begin{enumerate}[label=(\alph*), nosep, leftmargin=*]
    \item No ceiling, just the roof above
    \item A conventional ceiling
    \item A beamed ceiling
    \item A hanging ceiling
    \item A polythene cover as a ceiling
    \item No ceiling, the concrete slab
    \item Wooden ceiling
    \item Other
    \item Garden -- Not relevant
\end{enumerate}

\question{What is the main material used for the floor of this section of the household?}
\begin{enumerate}[label=(\alph*), nosep, leftmargin=*]
    \item Cement
    \item Teraso
    \item Tile
    \item Granite
    \item Wood (finished)
    \item Mud
    \item Wood
    \item Sand
    \item Concrete
    \item Other
    \item Garden -- Not relevant
\end{enumerate}

\question{How many doors are there in this part of the house that let in light and air from the outside? That is, how many doors are there that let in light and air into the room from an external environment, such as a garden, a balcony, or a veranda?}

\question{How many windows are there in this part of the house that let in light and air from the outside? That is, how many windows are there that let in light and air into the room from an external environment, such as a garden, a balcony, or a veranda?}

\question{Do you have curtains or blinds for the windows?}

\question{Does this section of the household have natural ventilation other than windows, to allow fresh air circulation? This could be a ventilation hole in the wall or an area with a grill on top of the windows, which is not covered by the window pane, etc.}

\question{How many light bulbs do you have in this section of the household?}

\question{And how many of these bulbs did you use in the last week?}

\question{How many fans do you have in this section of the household? Please include all types of fans such as ceiling fans, pedestal fans, table fans, exhaust fans, tower fans, etc.}

\question{How many ACs do you have in this section of the household?}

\question{What is the type of the light Bulb out of the following?}
\begin{enumerate}[label=(\alph*), nosep, leftmargin=*]
    \item Incandescent
    \item CFL
    \item LED
    \item Halogen
    \item Tube Light (conventional)
    \item Tube Light (LED)
    \item Flashlight
    \item Flood light
    \item Other
\end{enumerate}

\question{What is the wattage (W) of the light?}

\question{For how many hours was this light kept switched on last week during the daytime. (8 am -- 6 pm)?}

\question{For how many hours was this light kept switched on last week during nighttime. (6 pm -- 8 am)? \\
What is the type of the fan?}
\begin{enumerate}[label=(\alph*), nosep, leftmargin=*]
    \item Ceiling fan
    \item Wall fan
    \item Pedestal fan/ Stand fan
    \item Table fan
    \item Exhaust fan
    \item Tower fan
    \item Other
\end{enumerate}

\question{For how many hours was this fan kept switched on last week during the daytime. (8 am -- 6 pm)?}

\question{For how many hours was this fan kept switched on last week during nighttime. (6 pm -- 8 am)?}

\question{What is the type of air conditioning used in the room?}
\begin{enumerate}[label=(\alph*), nosep, leftmargin=*]
    \item Central AC (Only to your Household)
    \item Central AC (For the whole building. Relevant if B.5 = 5 or 6)
    \item Individual AC with two components
    \item Individual AC with one component
    \item Air Cooler
    \item Other
\end{enumerate}

\question{Is the AC an inverter AC?}

\question{Which of the following is true regarding air conditioners?}
\begin{enumerate}[label=(\alph*), nosep, leftmargin=*]
    \item The room can be fully closed and sealed, and there are no outside openings. When the AC is turned on, the cool air does not go out of the room.
    \item The room is fully closed. However, it is not fully sealed. Therefore, when the AC is on, the cool air may leak through the spaces that are not sealed, such as the space in-between the door and the door frame
    \item The room is not fully closed. There are spaces where the cool air can leak out
\end{enumerate}

\question{What is the BTU of the AC?}

\question{For how many hours was this AC kept switched on, last week, during daytime? (8 am -- 6 pm)?}

\question{For how many hours was this AC kept switched on last week, during nighttime. (6 pm -- 8 am)?}

\question{Are there any other parts of the household which are covered by the same electricity meter that were not covered above?}

\question{Can you please tell me which parts of the house you have missed mentioning earlier? Please mention only the parts you have missed, and to which electricity is supplied from the same electricity meter?}

\question{Are there any sections of the household that were there when we did the first interview, but have been demolished now?}

\question{Are there any constructions or renovations happening in this household currently?}

\question{Did the household had solar energy when the 1st interview was done?}

\question{Do you have a solar panel to generate electricity at your home?}

\question{You mentioned that you have solar panels to generate electricity when we did the first interview. Do you still have it?}

\question{Have you made any changes to the number of panels in the solar panel at your household after we did the first interview? i.e., have you increased or reduced the number of panels?}
\begin{enumerate}[label=(\alph*), nosep, leftmargin=*]
    \item Yes, we have increased the number of panels
    \item Yes, we have reduced the number of panels
    \item No, we have not made any changes to the number of panels
\end{enumerate}

\question{Is your solar system on-grid or off-grid?}

\question{Do you have an inverter for the solar energy generation system? An inverter is a device that converts direct current (DC) electricity, which is what a solar panel generates, to alternating current (AC) electricity, which the electrical grid uses.}

\question{Which of the following purposes do you use solar energy for inside your house?}
\begin{enumerate}[label=(\alph*), nosep, leftmargin=*]
    \item Solar Energy
    \item Bioenergy
    \item Mini Hydropower
    \item Wind Power
    \item Other
\end{enumerate}

\question{Are you aware of the number of units generated by the solar energy system in the last month? \\
How many units did the solar system generate last month?}

\question{What is the capacity of the solar system in your household?}

\question{Type of appliance and number of appliances of the type}

\question{Manufacturer/Brand}

\question{When was this purchased?}
\begin{enumerate}[label=(\alph*), nosep, leftmargin=*]
    \item Within the last 1 year.
    \item Within the last 1- 5 years.
    \item Within the last 5-10 years.
    \item More than 10 years ago.
\end{enumerate}

\question{Was it purchased as brand new or second-hand?}

\question{What is the door arrangement of the refrigerator?}
\begin{enumerate}[label=(\alph*), nosep, leftmargin=*]
    \item One door
    \item Two doors, freezer above the refrigerator
    \item Two doors, freezer below the refrigerator
    \item Three or more doors
    \item Other
\end{enumerate}

\question{Does your refrigerator require manual defrosting?}

\question{Does your refrigerator have the through-the-door ice facility?}

\question{What is the door arrangement of the freezer?}
\begin{enumerate}[label=(\alph*), nosep, leftmargin=*]
    \item Upright
    \item Chest
    \item Other
\end{enumerate}

\question{What is the defrosting type of the freezer?}
\begin{enumerate}[label=(\alph*), nosep, leftmargin=*]
    \item Manual
    \item Automatic
\end{enumerate}

\question{What is the type of your microwave?}
\begin{enumerate}[label=(\alph*), nosep, leftmargin=*]
    \item Solo
    \item Grill
    \item Convection
    \item Not aware
\end{enumerate}

\question{What is the type of your electric oven?}
\begin{enumerate}[label=(\alph*), nosep, leftmargin=*]
    \item Conventional
    \item Convectional
    \item Conventional with multifunction
    \item Not aware
\end{enumerate}

\question{What is the type of your electric cooktop?}
\begin{enumerate}[label=(\alph*), nosep, leftmargin=*]
    \item Solid plate
    \item Ceramic
    \item Induction
    \item Infra-red
    \item Cook top with a coil
\end{enumerate}

\question{What is the type of your washing machine?}
\begin{enumerate}[label=(\alph*), nosep, leftmargin=*]
    \item Fully automatic front-loading
    \item Fully automatic top-loading
    \item Semi-automatic Front loading
    \item Semi-automatic top-loading
\end{enumerate}

\question{Water temperature used for washing machine?}
\begin{enumerate}[label=(\alph*), nosep, leftmargin=*]
    \item Hot
    \item Warm
    \item Normal water
    \item Don't know
\end{enumerate}

\question{What is the type of your iron?}
\begin{enumerate}[label=(\alph*), nosep, leftmargin=*]
    \item Dry irons (with or without water steaming facility)
    \item Steam irons
    \item Other
\end{enumerate}

\question{What is the type of your TV?}
\begin{enumerate}[label=(\alph*), nosep, leftmargin=*]
    \item CRT
    \item LCD / LED
    \item OLED
    \item Plasma
\end{enumerate}

\question{How many times did you change the air filters in the last year?}
\begin{enumerate}[label=(\alph*), nosep, leftmargin=*]
    \item Once
    \item Twice
    \item Three times
    \item Four times
    \item More than four times
\end{enumerate}

\question{Size of the appliance?}

\question{Mention the length and the width of your Refrigerator.}

\question{Mention the length and the width of your Freezer}

\question{What is the plate size of your electric cooktop? By plate size, we mean the diameter of the cooktop.}

\question{Size of TV}

\question{Which of the following did you consider when purchasing the refrigerator? Please mention three most important attributes that you considered.}
\begin{enumerate}[label=(\alph*), nosep, leftmargin=*]
    \item Brand
    \item Energy Rating (wattage)
    \item Size of the refrigerator
    \item No. of doors in the refrigerator
    \item Defrosting type of the refrigerator
    \item Size of the freezer of the refrigerator
    \item Added Features
    \item Material
    \item Country of manufacture
    \item Color
    \item Safety
    \item Price
    \item Professional Reviews
    \item Other (specify)
\end{enumerate}

\question{Which of the following did you consider when purchasing the electric cooktop? Please mention three most important attributes that you considered.}
\begin{enumerate}[label=(\alph*), nosep, leftmargin=*]
    \item Brand
    \item Energy Rating
    \item Type of the cooktop (e.g., induction, infrared, etc.)
    \item Added Features
    \item Material
    \item Country of manufacture
    \item Color
    \item Safety
    \item Price
    \item Professional Reviews
    \item Other (specify)
\end{enumerate}

\question{Which of the following did you consider when purchasing the washing machine? Please mention three most important attributes that you considered.}
\begin{enumerate}[label=(\alph*), nosep, leftmargin=*]
    \item Brand
    \item Energy Rating (wattage)
    \item Type of washing machine (e.g., Fully automatic Front loading, fully automatic Top loading, semi-automatic Front loading, semi-automatic Top loading)
    \item Added Features
    \item Material
    \item Country of manufacture
    \item Color
    \item Safety
    \item Price
    \item Professional Reviews
    \item Other (specify)
\end{enumerate}

\question{Which of the following did you consider when purchasing the iron? Please mention three most important attributes that you considered.}
\begin{enumerate}[label=(\alph*), nosep, leftmargin=*]
    \item Brand
    \item Energy Rating (wattage)
    \item Added Features
    \item Material
    \item Type of iron (e.g., dry vs. steam)
    \item Country of manufacture
    \item Color
    \item Safety
    \item Price
    \item Professional Reviews
    \item Other (specify)
\end{enumerate}

\question{Which of the following did you consider when purchasing the AC? Please mention three most important attributes that you considered?}
\begin{enumerate}[label=(\alph*), nosep, leftmargin=*]
    \item Brand
    \item Energy Rating (wattage)
    \item BTU
    \item Added Features
    \item Material
    \item Country of manufacture
    \item Color
    \item Safety
    \item Price
    \item Professional Reviews
    \item Other (specify)
\end{enumerate}

\question{Which of the following best describes your behaviour of ironing clothes?}
\begin{enumerate}[label=(\alph*), nosep, leftmargin=*]
    \item We iron the clothes weekly.
    \item We iron our clothes twice a week.
    \item We iron the clothes daily.
    \item We iron when and where the need arises.
    \item We don't iron clothes.
\end{enumerate}

\question{Which of the following is true regarding drying the clothes that you wash under normal conditions?}
\begin{enumerate}[label=(\alph*), nosep, leftmargin=*]
    \item We don't use the dryer / in-built dryer in the washing machine to dry our clothes, we dry them out naturally (sunlight)
    \item If we use the washing machine, we use the built-in dryer to dry our clothes.
    \item We always use the dryer / in-built dryer in the washing machine to dry our clothes, irrespective of how they were washed.
\end{enumerate}

\question{Which of the following best describes your behaviour regarding night lighting, considering the whole house?}
\begin{enumerate}[label=(\alph*), nosep, leftmargin=*]
    \item We don't keep any of the lights on when sleeping.
    \item We keep one or two night lights on when sleeping.
    \item We keep more than two lights on when sleeping.
\end{enumerate}

\question{How often do you and your household members switch off unnecessary lights when leaving a room?}
\begin{enumerate}[label=(\alph*), nosep, leftmargin=*]
    \item Always
    \item Sometimes
    \item Never
\end{enumerate}

\question{How conscious are you and your household members about reducing the number of times you open and close the refrigerator?}
\begin{enumerate}[label=(\alph*), nosep, leftmargin=*]
    \item Very conscious
    \item Somewhat conscious
    \item Not conscious at all
\end{enumerate}

\question{When buying appliances, do you and your household members look at the wattage (or the energy rating) for energy saving purposes?}

\question{What is the total expenditure of your household in the last month? Please consider the spending by all members in your household. Please include all types of expenses, such as food items that you or your household members bought such as  rice, coconut, chillies, sugar, prepared food that you'll bought from outside, you and your household members' spending on personal care products such as soaps, shampoo, toothpaste etc., expenses on laundry products such as washing powder, washing soap, expenses on travelling, on education such as school fees or tuition fees etc., bill payments such as electricity bill, water bill phone bill etc., your spending on clothes etc.}

\question{As you know, we are doing this study regarding the electricity consumption. Therefore, can you please tell me what type of electricity meter you have at home? Is it a smart meter or a non-smart meter?}

\section{Survey Wave 3}
\setcounter{qcounter}{0}

\question{Are you the same respondent as in wave 1 and/or wave 2, or is it a different member of the household?}
\begin{enumerate}[label=(\alph*), nosep, leftmargin=*]
    \item Same respondent as in wave 1 / wave 2
    \item A different respondent compared to wave 1 / wave 2
\end{enumerate}

\question{Name of respondent}

\question{Are you able to accurately tell me about the details related to the household?}

\question{Do you consent to participate in this study and to provide your data to LIRNEasia?}

\question{May we audio record some parts of this interview for quality checking purposes?}

\question{Have you obtained any new electricity meter/s since we did the last survey?}

\question{Please tell me how many electricity meters are there in your house now?}

\question{Do you or any member of your household own this house, or are you living on rent?}
\begin{enumerate}[label=(\alph*), nosep, leftmargin=*]
    \item Yes, I or a household member owns it.
    \item No, I am living on rent, and I or a household member pays the rent.
    \item No, I am living on rent, and the employer pays the rent.
    \item No, I or any household member does not own or rent this household. We occupy this household without any payment of rent.
\end{enumerate}

\question{Do you occupy any of the following people in your house? By house, we mean the area covered by your electricity meter.}
\begin{enumerate}[label=(\alph*), nosep, leftmargin=*]
    \item Renters/boarders who are living in your annexe or any other attached place, maintaining separate living conditions but sharing the same electricity meter.
    \item Boarders who live in your house using a room/s that are attached to your living conditions.
    \item I don't occupy any of the above.
\end{enumerate}

\question{You said that renters/boarders are living in this house. Can you please tell me whether they were living in this house when we did the previous interviews, or did they start living here afterwards?}
\begin{enumerate}[label=(\alph*), nosep, leftmargin=*]
    \item They were living in this household from the time when the 2nd interview was done.
    \item They started living in this household after the 2nd interview was done.
    \item Some were there when the second interview was done, but some came after the second interview.
\end{enumerate}

\question{Since when did the new renters/boarder (s) start living in this household?}
\begin{enumerate}[label=(\alph*), nosep, leftmargin=*]
    \item December 2023
    \item January 2024
    \item February 2024
    \item March 2024
    \item April 2024
    \item May 2024
    \item June 2024
    \item July 2024
    \item August 2024
    \item September 2024
    \item October 2024
    \item November 2024
    \item December 2024
    \item Other (specify)
\end{enumerate}

\question{Which of the following best describes your awareness of the electricity consumption of renters/boarders? Are you aware of details such as the appliances they use and the number of hours they use each appliance, the times they keep the lights and fans switched on, etc.?}
\begin{enumerate}[label=(\alph*), nosep, leftmargin=*]
    \item I know all the details about the electricity consumption of the renters/ boarders; i.e., the appliances they use and the number of hours they use each appliance, the times they keep the lights and fans switched on, etc.
    \item I know some details about the electricity consumption of the renters/ boarders; i.e., the appliances they use and the number of hours they use each appliance, the times they keep the lights and fans switched on, etc.
    \item I do not know any details about the electricity consumption of the renters/ boarders; i.e., the appliances they use and the number of hours they use each appliance, the times they keep the lights and fans switched on, etc.
\end{enumerate}

\question{Has there been any structural changes to your household after we did the 2nd interview with you? By household, we mean the area covered by your electricity meter. i.e., have there been any additions to the house, such as adding a new part to the household like a room, etc., or completing a section which was being built at the time we did the first interview, or added a new floor, etc.? or, have there been any parts of the household which were dismantled?}
\begin{enumerate}[label=(\alph*), nosep, leftmargin=*]
    \item Yes, new parts were added only
    \item Yes, some parts were dismantled only
    \item Yes, some new parts were added to the household, and some parts were dismantled
    \item No structural changes were done to the household
\end{enumerate}

\question{Were there any renovations done after we did the 2nd interview with you?}

\question{Which of these best describes your house now?}
\begin{enumerate}[label=(\alph*), nosep, leftmargin=*]
    \item Single House - Single Floor
    \item Single House -Double Floor
    \item Single House -- More than 2 floors
    \item Attached house / Annex
    \item Flat
    \item Condominium/ Luxury apartments
    \item Twin houses
    \item Line room/row house
    \item Slum / Shanty
    \item Other
\end{enumerate}

\question{How many storeys does your house have?}

\question{Can you please tell me what the floor area of this household is now? i.e., how many square feet does this household cover? Please remember, when we say household, what we mean is the parts of the household that are covered by the electricity meter.}

\question{During the last two interviews we had with you, you mentioned that there are \_\_\_\_ members living in the household. Can you please tell me whether there are any changes in the composition of members living in the household? i.e., whether these members are still living in this household or whether they have moved out and hence, are not staying here anymore? If they are temporarily living abroad or living in some other part of Sri Lanka, please consider them as living in the household. Also, if there are any boarders or those who are living on rent, or if you have any maids/ housekeepers staying in this household, please include them as well.}

\question{Apart from the above-mentioned members, are there any others who are currently living in the household? It could be someone who has come back from abroad, or any new additions? It could also be a boarder or a domestic help who has started living in the household after (MENTION MONTH WHEN FIELDWORK WAS CONDUCTED FOR WAVE 2)? If so, how many such members are there?}

\question{Name of the member}

\question{Relationship to the head of the household.}
\begin{enumerate}[label=(\alph*), nosep, leftmargin=*]
    \item Head of the household
    \item Wife/Husband
    \item Son/daughter
    \item Son-in-law/Daughter in law
    \item Parents of the head of the Household/ spouse
    \item Grandson/ Granddaughter
    \item Other relative
    \item Domestic servant/driver/watcher
    \item Boarder
    \item Other
\end{enumerate}

\question{Gender}
\begin{enumerate}[label=(\alph*), nosep, leftmargin=*]
    \item Male
    \item Female
    \item Other
\end{enumerate}

\question{Year of Birth}

\question{Age}

\question{Ethnicity}
\begin{enumerate}[label=(\alph*), nosep, leftmargin=*]
    \item Sinhala
    \item Sri Lankan Tamil
    \item Indian Tamil
    \item Sri Lankan Moor/Muslim
    \item Burgher
    \item Malay
    \item Other
\end{enumerate}

\question{Religion}
\begin{enumerate}[label=(\alph*), nosep, leftmargin=*]
    \item Buddhism
    \item Hinduism
    \item Islam
    \item Roman Catholicism
    \item Other Christian denominations
    \item No religion
    \item Other
\end{enumerate}

\question{Marital status}
\begin{enumerate}[label=(\alph*), nosep, leftmargin=*]
    \item Never married
    \item Currently married (registered)
    \item Currently married (customary)
    \item Widowed
    \item Divorced
    \item Legally separated
    \item Separated (not legally)
    \item Not married but lives as a family
    \item Other
\end{enumerate}

\question{Current attendance at school or any other educational institution. i.e., is he/she currently doing any type of studying? This could be a degree, an MBA, etc., or following a course like CIMA, CIM, or a vocational training course?}
\begin{enumerate}[label=(\alph*), nosep, leftmargin=*]
    \item Preschool
    \item School
    \item University
    \item Other educational institution
    \item Vocational/Technical Institution
    \item Pending results G.C.E. (O.L / A.L)
    \item Still a toddler
    \item Does not attend
\end{enumerate}

\question{Highest level of education}
\begin{enumerate}[label=(\alph*), nosep, leftmargin=*]
    \item Studying/Studied Grade 1
    \item Passed Grade 1
    \item Passed Grade 2
    \item Passed Grade 3
    \item Passed Grade 4
    \item Passed Grade 5
    \item Passed Grade 6
    \item Passed Grade 7
    \item Passed Grade 8
    \item Passed Grade 9
    \item Passed Grade 10
    \item Passed G.C.E. (O/L)
    \item Passed Grade 12
    \item Passed G.C.E. (A/L) or equivalent
    \item Passed GAQ (General Arts Qualifying (External) Examination) / GSQ (General Science Qualifying Examination)
    \item Passed Degree / Diploma
    \item Passed Post Graduate Degree / Diploma
    \item PhD
    \item Special Education learning / learnt
    \item No Schooling
\end{enumerate}

\question{Main activity usually engaged in}
\begin{enumerate}[label=(\alph*), nosep, leftmargin=*]
    \item Engaged in economic activity / currently employed/engaged in own business
    \item Retired and obtaining government/semi-government pension payment and is currently engaged in economic activity (employed elsewhere other than the place where he/she is receiving the pension from / engaged in own business)
    \item Seeking and available to work
    \item Retired - Obtaining government/semi-government pension payment and currently not engaged in economic activity (not employed elsewhere or not engaged in any own business)
    \item Retired from the private/semi-government sector and does not receive any pension payment.
    \item Received other pension payments.
    \item Household activities
    \item Student
    \item Too old / Disabled/unable to work.
    \item Other
\end{enumerate}

\question{Main occupation}
\begin{enumerate}[label=(\alph*), nosep, leftmargin=*]
    \item Legislator, senior official, and manager
    \item Professional
    \item Technician and associate professional
    \item Clerk
    \item Service worker, and shop and market sales worker
    \item Skilled agricultural and fishery worker
    \item Craft and related worker
    \item Plant and machine operator and assembler
    \item Elementary occupation
    \item Related to forces
    \item No occupation
\end{enumerate}

\question{Daily wage owner or not}

\question{Employment status of the main occupation}
\begin{enumerate}[label=(\alph*), nosep, leftmargin=*]
    \item Government employee
    \item Semi-government employee
    \item Private sector employee
    \item Employer
    \item Own account worker
    \item Contributing family worker
\end{enumerate}

\question{The electricity consumption can vary based on the time that people stay at home. Therefore, we would like to know the number of hours spent at home during the last week for each member.}

\question{During the last week, did this member go out of the home for employment purposes?}
\begin{enumerate}[label=(\alph*), nosep, leftmargin=*]
    \item Yes, went daily during working days.
    \item Yes, went on most of the days.
    \item No, worked from home.
\end{enumerate}

\question{I would also like to know a few details about the other members of the household. \\ Name of the member}

\question{Was there any change in the main activity usually engaged in? It could be a change in jobs, a promotion, etc., or started working full-time or part-time, or you started studying full-time, stopped working/retired, etc.}

\question{Main activity usually engaged in?}
\begin{enumerate}[label=(\alph*), nosep, leftmargin=*]
    \item Engaged in economic activity / currently employed/engaged in own business
    \item Retired and obtaining government/semi-government pension payment and is currently engaged in economic activity (employed elsewhere other than the place where he/she is receiving the pension from / engaged in own business)
    \item Seeking and available to work
    \item Retired - Obtaining government/semi-government pension payment and currently not engaged in economic activity (not employed elsewhere or not engaged in any own business)
    \item Retired from the private/semi-government sector and does not receive any pension payment.
    \item Received other pension payments.
    \item Household activities
    \item Student
    \item Too old / Disabled/unable to work.
    \item Other
\end{enumerate}

\question{Main occupation}
\begin{enumerate}[label=(\alph*), nosep, leftmargin=*]
    \item Legislator, senior official, and manager
    \item Professional
    \item Technician and associate professional
    \item Clerk
    \item Service worker, and shop and market sales worker
    \item Skilled agricultural and fishery worker
    \item Craft and related worker
    \item Plant and machine operator and assembler
    \item Elementary occupation
    \item Related to forces
    \item No occupation
\end{enumerate}

\question{Daily wage owner?}

\question{Employment status of main occupation.}
\begin{enumerate}[label=(\alph*), nosep, leftmargin=*]
    \item Government employee
    \item Semi-government employee
    \item Private sector employee
    \item Employer
    \item Own account worker
    \item Contributing family worker
\end{enumerate}

\question{The electricity consumption can vary based on the time that people stay at home. Therefore, we would like to know the number of hours spent at home during the last week for each member.}

\question{You mentioned that renters or boarders are living in this household who share the same electricity meter. Which of the following best describes how you charge them for electricity?}
\begin{enumerate}[label=(\alph*), nosep, leftmargin=*]
    \item You charge a fixed amount every month for electricity.
    \item You charge an amount for electricity depending on the variance of the bill.
    \item You don't charge a specific amount for electricity, but charge a fixed amount for all the utilities, such as electricity, water, etc.
    \item You don't charge a specific amount for electricity, but charge a varied amount for all the utilities, such as electricity, water, etc. The amount charged varied based on the utility bills
    \item You don't charge them for electricity consumption
\end{enumerate}

\question{You mentioned that you are living on rent. Which of the following best describes the payment for your electricity consumption?}
\begin{enumerate}[label=(\alph*), nosep, leftmargin=*]
    \item You pay the full amount of the electricity bill.
    \item You pay a fixed amount to the owner every month for electricity.
    \item You pay a varied amount to the owner every month for electricity. The amount paid varies depending on the variance of the bill.
    \item You don't pay a specific amount for electricity, but pay a fixed amount for all the utilities, such as electricity, water, etc.
    \item You don't pay a specific amount for electricity, but pay a varied amount for all the utilities, such as electricity, water, etc. The amount paid varies depending on the variance of the utility bills.
    \item You don't pay the owner for electricity consumption
\end{enumerate}

\question{Is there any business activity that is being carried out in any part of this household, such as a shop, a communication, your own office, etc., for which electricity is used from the same meter?}

\question{What type of business activity is being carried out in this household?}
\begin{enumerate}[label=(\alph*), nosep, leftmargin=*]
    \item A shop
    \item A communication
    \item Other (specify)
\end{enumerate}

\question{You mentioned you had \_\_\_ (MENTION THE NUMBER OF APPLIANCE) \_\_\_\_ (MENTION THE CORRESPONDING TYPE OF APPLIANCE). Do you still have the same number of \_\_\_(MENTION APPLIANCE TYPE) in working condition?}

\question{No of appliances of the type}

\question{Duration of usage in hours during the last week}

\question{Do you have any of these appliances in your household in working condition, which you have used at least once? If you have boarders or those who are living on rent in your household, please include the electrical appliances used by them as well.}
\begin{enumerate}[leftmargin=*]
    \item Refrigerator
    \item Separate Freezer
    \item Mini Bar
    \item Microwave
    \item Electric Oven
    \item Electric cook tops (induction cookers, Infra-red cookers, hot plates)
    \item Electrical exhaust fan fitted above the oven or the hot plate
    \item Electric Blender
    \item Electric grinder
    \item Electric mixer/beater
    \item Electric food processor
    \item Rice cooker
    \item Electric Kettle
    \item Electric water heater to heat water for drinking purposes
    \item Toaster / Sandwich toaster
    \item Waffle maker
    \item Electric pressure cooker
    \item Air fryer
    \item Coffee maker
    \item Electric grill
    \item Electric coconut scraper
    \item Electric Water filter/water dispenser
    \item Dishwasher
    \item Washing Machine
    \item Clothes dryer
    \item Electric Iron, including electric steam iron
    \item Electric Vacuum Cleaner
    \item Electric floor polisher
    \item Electric water gun (used to wash cars, etc.)
    \item Electric Lawn mower
    \item TV
    \item TV antenna
    \item Dialog TV / Peo TV / Satellite TV box
    \item Radio
    \item Bluetooth Speakers
    \item DVD / VCD
    \item Gaming console/PlayStation
    \item Sound systems (Subwoofer)/Stereo
    \item Camera (that needs to be charged using electricity)
    \item Home theater system
    \item Electric musical Instruments (ex., electric organ, electric guitar, etc.)
    \item Computers
    \item Laptops
    \item Routers
    \item Mobile phone - Smartphones
    \item Mobile phone - Feature phones
    \item Mobile phone - Basic phones
    \item Fixed phones
    \item Power banks
    \item Printer
    \item Scanner
    \item Fax Machines
    \item Photo Copiers
    \item Hair Dryer
    \item Hair Iron/Hair Curlers
    \item Electric Shavers
    \item Electric exercise machines
    \item Roller door
    \item CCTV camera systems
    \item Electric Alarm system
    \item Other electric security systems
    \item Electric bell
    \item Electric vehicles (four wheelers)
    \item Electric vehicles (two wheelers)
    \item Electric vehicles (three wheelers)
    \item Electric Water Pump
    \item Electric Water pump
    \item Electric Sewing machine
    \item Oxygen filter for fish tank
    \item Geyser / Hot water systems for bathrooms, which operate from electricity
    \item Hot tub
    \item Electric Fountain / decorative waterfall
    \item Emergency Light / rechargeable torches
    \item Electric heater (to control room temperature)
    \item Humidifier
    \item Toys rechargeable batteries
    \item Other (specify)
\end{enumerate}

\question{No of appliances of each type}

\question{Duration of usage in hours during the last week for each type}

\question{Are there any changes to the number of \_\_\_\_ (room type)?}
\begin{enumerate}
    \item  Living room
    \item Bedrooms
    \item Kitchen and/or pantry/dining room
    \item Bathroom and/or toilets
    \item Storage room
    \item Other rooms/parts in the house, including passages, verandas, balconies, staircase, etc.
    \item Garage
    \item Security room
    \item Garden
    \item Other rooms outside the house
\end{enumerate}

\question{New no. of rooms for each type}

\textbf{\\Repeat Q55 to Q81 for each room in the house}

\question{For what purposes do you mainly use this room?}
\begin{enumerate}[label=(\alph*), nosep, leftmargin=*]
    \item Living room
    \item Bedrooms
    \item Kitchen and/ or pantry
    \item Bathroom and/or toilets
    \item Storage room
    \item Gaming room
    \item Servant's Room
    \item Visitors' room
    \item Passage
    \item Veranda
    \item Balcony
    \item Staircase
    \item Garage
    \item Security Room
    \item Garden
    \item Other rooms outside the house
    \item Study room /Office room
    \item Other
\end{enumerate}

\question{Was this section a part of the household when we did the second interview in (MENTION THE MONTH WHEN FIELDWORK WAS CONDUCTED FOR WAVE 2)? And were there any changes made for this section of the household?}
\begin{enumerate}[label=(\alph*), nosep, leftmargin=*]
    \item Yes, this was a part of the house when the second interview was done, and no changes were made to this section of the household
    \item Yes, this was a part of the house when the second interview was done, but there were some changes made to this section of the household
    \item No, this section was built after the second interview
\end{enumerate}

\question{In which storey is this room located? (Define ground floor by zero)}

\question{What is the main material used for the roof of this section of the household?}
\begin{enumerate}[label=(\alph*), nosep, leftmargin=*]
    \item Tile
    \item Asbestos
    \item Concrete
    \item Metal Sheet
    \item Takaran
    \item Cadjun/Palmyra/Straw
    \item Tent
    \item Plastic sheets
    \item Other
    \item Garden -- Not relevant
\end{enumerate}

\question{What type of ceiling does this section of the household have?}
\begin{enumerate}[label=(\alph*), nosep, leftmargin=*]
    \item No ceiling, just the roof above
    \item A conventional ceiling
    \item A beamed ceiling
    \item A hanging ceiling
    \item A polythene cover as a ceiling
    \item No ceiling, the concrete slab
    \item Wooden ceiling
    \item Other
    \item Garden -- Not relevant
\end{enumerate}

\question{What is the main material used for the floor of this section of the household?}
\begin{enumerate}[label=(\alph*), nosep, leftmargin=*]
    \item Cement
    \item Teraso
    \item Tile
    \item Granite
    \item Wood (finished)
    \item Mud
    \item Wood
    \item Sand
    \item Concrete
    \item Other
    \item Garden -- Not relevant
\end{enumerate}

\question{How many doors are there in this part of the house that let in light and air from the outside? That is, how many doors are there that let in light and air into the room from an external environment, such as a garden, a balcony, or a veranda?}

\question{How many windows are there in this part of the house that let in light and air from the outside? That is, how many windows are there that let in light and air into the room from an external environment, such as a garden, a balcony, or a veranda?}

\question{Do you have curtains or blinds for the windows?}

\question{Does this section of the household have natural ventilation other than windows, to allow fresh air circulation? This could be a ventilation hole in the wall or an area with a grill on top of the windows, which is not covered by the window pane, etc.}

\question{How many light bulbs do you have in this section of the household? If there are lamp shades, etc., which have more than one bulb, but all bulbs switch on with one common switch, please count as one light. Also, if there are table lamps, bedside lamps, lamp shades, etc., in this part of the household, please include them as well.}

\question{And how many of these bulbs did you use in the last week?}

\textbf{\\Repeat Q67 to Q70 for each light bulb in the room}

\question{What is the type of the light Bulb out of the following?}
\begin{enumerate}[label=(\alph*), nosep, leftmargin=*]
    \item Incandescent
    \item CFL
    \item LED
    \item Halogen
    \item Tube Light (conventional)
    \item Tube Light (LED)
    \item Flashlight
    \item Flood light
    \item Other
\end{enumerate}

\question{What is the wattage (W) of the light?}

\question{For how many hours was this light kept switched on last week during daytime. (8 am -- 6 pm)?}

\question{For how many hours was this light kept switched on, during last week, during nighttime. (6 pm -- 8 am)?}

\textbf{\\Repeat Q72 to Q74 for each fan in the room}

\question{How many fans do you have in this section of the household? Please include all types of fans such as ceiling fans, pedestal fans, table fans, exhaust fans, tower fans, etc.}

\question{What is the type of the fan?}
\begin{enumerate}[label=(\alph*), nosep, leftmargin=*]
    \item Ceiling fan
    \item Wall fan
    \item Pedestal fan
    \item Table fan
    \item Exhaust fan
    \item Tower fan
    \item Other
\end{enumerate}

\question{For how many hours was this fan kept switched on last week during daytime. (8 am -- 6 pm)}

\question{For how many hours was this fan kept switched on last week during nighttime. (6 pm -- 8 am)?}

\question{How many ACs do you have in this section of the household?}

\question{What is the type of air conditioning used in the room?}
\begin{enumerate}[label=(\alph*), nosep, leftmargin=*]
    \item Central AC (Only to your Household)
    \item Central AC
    \item Individual AC with two components
    \item Individual AC with one component
    \item Air Cooler
    \item Other
\end{enumerate}

\question{Is the AC an inverter AC? Repeat for each AC in the room.}

\question{Which of the following is true regarding air conditioners?}
\begin{enumerate}[label=(\alph*), nosep, leftmargin=*]
    \item The room can be fully closed and sealed, and there are no outside openings. When the AC is turned on, the cool air does not leave the room.
    \item The room is fully closed. However, it is not fully sealed. Therefore, when the AC is on, the cool air may leak through the spaces that are not sealed, such as the space in-between the door and the door frame
    \item The room is not fully closed. There are spaces where the cool air can leak out.
\end{enumerate}

\question{What is the BTU of the AC? Repeat for each AC in the room.}

\question{For how many hours was this AC kept switched on, last week, during daytime? (8 am -- 6 pm)? Repeat for each AC in the room.}

\question{For how many hours was this AC kept switched on last week, during nighttime. (6 pm -- 8 am)? Repeat for each AC in the room.}

\question{Are there any other parts of the household which are covered by the same electricity meter that were not covered above?}

\question{Can you please tell me which parts of the house you have missed mentioning earlier? Please mention only the parts you have missed, and to which electricity is supplied from the same electricity meter?}
\begin{enumerate}
    \item Living room
    \item Bedrooms
    \item Kitchen and/ or pantry/dining room
    \item Bathroom and/or toilets
    \item Storage room
    \item Other rooms/parts in the house, including passages, verandas, balconies, staircase, etc.
    \item Garage
    \item Security Room
    \item Garden (including the wall around the garden, if any)
    \item Other rooms outside the house
\end{enumerate}

\question{Are there any sections of the household that were there when we did the second interview, but have been demolished now?}

\question{Can you please tell me which parts of the house were demolished after we did the second interview? And how many such parts were demolished?}
\begin{enumerate}
    \item Living room
    \item Bedrooms
    \item Kitchen and/ or pantry/dining room
    \item Bathroom and/or toilets
    \item Storage room
    \item Other rooms/parts in the house, including passages, verandas, balconies, staircase, etc.
    \item Garage
    \item Security Room
    \item Garden (including the wall around the garden, if any)
    \item Other rooms outside the house
\end{enumerate}

\question{Are there any constructions or renovations happening in this household currently?}

\question{Does your household have a backup generator to generate electricity if needed?}

\question{Did the household had solar energy when the 2nd interview was done?}

\question{Do you have a solar panel to generate electricity at your home?}

\question{You mentioned that you have solar panels to generate electricity when we did the second interview. Do you still have it?}

\question{Have you made any changes to the number of panels in the solar panel at your household after we did the second interview? i.e., have you increased or reduced the number of panels?}
\begin{enumerate}[label=(\alph*), nosep, leftmargin=*]
    \item Yes, we have increased the number of panels
    \item Yes, we have reduced the number of panels
    \item No, we have not made any changes to the number of panels
\end{enumerate}

\question{Is your solar system on-grid or off-grid?}
\begin{enumerate}[label=(\alph*), nosep, leftmargin=*]
    \item On-grid
    \item Off-grid
    \item Don't know /Can't say
\end{enumerate}

\question{Do you have an inverter for the solar energy generation system? An inverter is a device that converts direct current (DC) electricity, which is what a solar panel generates, to alternating current (AC) electricity, which the electrical grid uses}

\question{Which of the following purposes do you use solar energy for inside your house?}
\begin{enumerate}[label=(\alph*), nosep, leftmargin=*]
    \item Water heating
    \item Cooking
    \item Outdoor lighting
    \item Car charging
    \item Agriculture equipment and systems (irrigation systems, etc)
    \item All the above
    \item Other
\end{enumerate}

\question{Are you aware of the number of units generated by the solar energy system in the last month?}

\question{How many units did the solar system generate last month?}

\question{What is the capacity of the solar system in your household?}

\question{Have you and your household members consciously reduced or stopped using any electrical
appliance, i.e., consciously switched off or stopped using electrical devices such as fans, electric
rice cookers, kettles, ovens, etc.?}
\begin{enumerate}[label=(\alph*), nosep, leftmargin=*]
    \item Yes, always consciously reduce/ stop using some appliances
    \item Yes, sometimes consciously reduce/ stop using some appliances
    \item No, never consciously reduce/ stop using some appliances
\end{enumerate}

\question{Which appliances/devices have you and your household members reduced or stopped
using consciously?}

\question{Do you or your household members switch off the refrigerator as a habit at specific times of the day? This does not include switching off the refrigerator when there is lightning, etc.}

\question{To what extent do you agree with the statement ``We usually replace the old electric appliances
and buy new ones without waiting till the appliance is unusable''?}
\begin{enumerate}[label=(\alph*), nosep, leftmargin=*]
    \item Very much disagree
    \item Somewhat disagree
    \item Neither agree nor disagree
    \item Somewhat agree
    \item Very much agree
\end{enumerate}

\question{Have you heard of ``Time of use metering''?}

\question{Have you or your household used ``Time of use metering''?}

\question{Has the use of ``Time of use metering'' reduced your electricity bill considerably? i.e., reduced your electricity bill by at least 20\%?}

\question{Why have you not used ``Time of use metering''?}
\begin{enumerate}[label=(\alph*), nosep, leftmargin=*]
    \item I wanted to apply, but couldn't do it yet
    \item Don't think there is any considerable reduction in the electricity bill by using ``Time of use
    monitoring''
    \item I don't know how to apply for it
    \item Other (specify)
\end{enumerate}

\question{The ``Time of use metering'' is a way of measuring electricity consumption separately according to the different times of day. The day is divided into 3 periods: peak, daytime, and off-peak, and
electricity usage is charged at different rates for the three periods, with a higher rate for the peak,
a medium rate for the daytime, and a lower rate for the off-peak time. In addition to the charge based on the time period, there is also a fixed charge per month.\\
If this type of electricity billing were available, would you use it?}

\question{You mentioned that if ``Time of use metering'' were available, you would use it. Will you be
able to shift your electrical appliance usage to off-peak times to gain a considerable
reduction (at least a 20\% reduction) in your electricity bill?}

\question{Do you have a target number of units or a target bill value for monthly electricity consumption for your household?}

\question{What is your target type?}
\begin{enumerate}[label=(\alph*), nosep, leftmargin=*]
    \item Number of units
    \item Bill value
    \item Both
\end{enumerate}

\question{What was the target maximum number of units for the last month?}

\question{What was the target bill value for the last month?}

\question{Does any household member view smart meter interval data?}

\question{Which of the following can most accurately describe your knowledge in electricity bill calculation?}
\begin{enumerate}[label=(\alph*), nosep, leftmargin=*]
    \item I am aware of how the bill is calculated.
    \item I'm not aware of how the bill is calculated, but a household member is aware
    \item Neither I nor my household members are aware of how the bill is calculated
\end{enumerate}

\question{Which of the below is true regarding your normal bill payment practice (example given when the
bill is Rs 4566)?}
\begin{enumerate}[label=(\alph*), nosep, leftmargin=*]
    \item We try to pay the exact amount in the bill (ex: 4566)
    \item We try to pay a rounded-off amount, which is mostly lower than the bill (ex: 4500)
    \item We try to pay a rounded off amount which will cover the whole bill amount (ex: 4600)
    \item We try to pay a portion of the bill amount that is possible for us (ex: 3500)
    \item We don't pay the bill regularly/ every month.
    \item We live on rent and do not pay directly to LECO.
\end{enumerate}

\question{Have you experienced any disconnection in electricity supply due to late payment during the last year?}

\question{Nearly how many red notices did you receive last year from LECO? A red notice is issued for failing to pay outstanding bills during a certain period.}

\question{How much did your family spend last month? Consider the spending of all members of your
household. Please include all types of expenses, such as food items that you or your household
members bought such as rice, coconut, chilies, sugar, and prepared food that you bought from
outside, you and your household members' spending on personal care products such as soaps,
shampoo, toothpaste, etc., expenses on laundry products such as washing powder, washing soap,
expenses on travel, education, such as school fees or tuition fees, etc., your spending on
clothes, etc.}

\question{As you know, we are doing this study regarding the electricity consumption. Therefore, can you
please tell me what type of electricity meter you have at home? Is it a smart meter or a non-
smart meter?}

\supplementheading{C}{DATASHEET}
\definecolor{ieeedata}{cmyk}{0.34,0.97,0.59,0.28}

\newcommand{\dssectionheader}[1]{%
   \noindent\framebox[\columnwidth]{%
      {\fontfamily{phv}\selectfont \textbf{\textcolor{ieeedata}{#1}}}
   }
}

\newcommand{\dsquestion}[1]{%
    {\noindent {\textbf{#1}}}
}

\newcommand{\dsquestionex}[2]{%
        {\noindent {\textbf{#1 \textit{#2}} }} 

}

\newcommand{\dsanswer}[1]{%
   {\noindent #1 \medskip \medskip}
}

\dssectionheader{Motivation}

\dsquestionex{For what purpose was the dataset created?}{Was there a specific task in mind? Was there a specific gap that needed to be filled? Please provide a description.}

\dsanswer{The dataset was developed to address a gap in the availability of granular data necessary for designing AI-based demand-side management strategies. At the time of creation, no comparable dataset existed in Sri Lanka. Even globally, few datasets offer the same level of integration between high-resolution electricity consumption data and detailed temporal and contextual information on the demographic, social, and physical characteristics of households.
}

\dsquestion{Who created this dataset (e.g., which team, research group) and on behalf of which entity (e.g., company, institution, organization)?}

\dsanswer{The dataset was developed by the Data, Algorithms, and Policy Team at LIRNEasia, a Sri Lanka-based public policy think tank operating across the Asia-Pacific region. LIRNEasia has extensive experience in conducting large-scale surveys, advancing data science research, and working with transactional data to inform policy and promote the public good. This project was carried out in collaboration with the Lanka Electricity Company (LECO), one of Sri Lanka's electricity distribution licensees, which provided access to de-identified electricity consumption data and supported the design and implementation of the longitudinal survey.
}

\dsquestionex{Who funded the creation of the dataset?}{If there is an associated grant, please provide the name of the grantor and the grant name and number.}

\dsanswer{ The creation of this dataset was funded by Lacuna Fund, a collaborative initiative that provides resources for developing labeled datasets to address urgent problems in low- and middle-income contexts globally.
\begin{enumerate}
    \item \textbf{Grant Name}: 20-Month Dataset on Household Electricity Consumption and its Drivers, Collected via Meter Readings, and Longitudinal Survey
    \item \textbf{Grant Number}: Grantee \#55 (19497.59)
\end{enumerate}
}

\bigskip
\dssectionheader{Composition}

\dsquestionex{What do the instances that comprise the dataset represent (e.g., documents, photos, people, countries)?}{ Are there multiple types of instances (e.g., movies, users, and ratings; people and interactions between them; nodes and edges)? Please provide a description.}

\dsanswer{The instances in the dataset represent households}

\dsquestion{How many instances are there in total (of each type, if appropriate)?}

\dsanswer{4063 instances}

\dsquestionex{Does the dataset contain all possible instances or is it a sample (not necessarily random) of instances from a larger set?}{ If the dataset is a sample, then what is the larger set? Is the sample representative of the larger set (e.g., geographic coverage)? If so, please describe how this representativeness was validated/verified. If it is not representative of the larger set, please describe why not (e.g., to cover a more diverse range of instances, because instances were withheld or unavailable).}

\dsanswer{The dataset is a sample, not a complete enumeration. It was drawn from the customer database of the Lanka Electricity Company (LECO), which serves approximately 500,000 households along the western coastal belt of Sri Lanka. The sample comprises 4,063 households---1,438 smart-metered and 2,625 non-smart-metered---randomly selected to reflect the broader LECO customer base. Sampling weights were applied based on the number of households served in each regional service area to ensure geographic representativeness across LECO's coverage area. While the dataset is representative of LECO customers, it does not extend to households served by other electricity providers in Sri Lanka.}

\dsquestionex{What data does each instance consist of? ``Raw'' data (e.g., unprocessed text or images) or features?}{In either case, please provide a description.}

\dsanswer{Each instance in the dataset consists of only "Raw" data. The raw data includes electricity consumption records from smart meters, collected at 15-minute and 6-hour intervals, as well as monthly readings for non-smart-metered households. Smart meter data also includes voltage, current, and frequency measurements. Additionally, the dataset contains longitudinal survey responses covering household demographics, housing characteristics, appliance ownership, energy security, and attitudes toward electricity use.}

\dsquestionex{Is there a label or target associated with each instance?}{If so, please provide a description.}

\dsanswer{The dataset does not have a single predefined label but includes multiple variables that can serve as targets depending on the research question. For electricity consumption data, potential targets include smart meter readings (15-minute or 6-hour intervals), monthly consumption totals, and voltage, current, and frequency measurements. For survey data, targets may include demographic attributes, appliance usage patterns, and self-reported energy security and attitudes. These variables support predictive modeling tasks such as forecasting energy demand, identifying inefficient appliances, and analyzing demographic impacts on electricity use.
}

\dsquestionex{Is any information missing from individual instances?}{If so, please provide a description, explaining why this information is missing (e.g. because it was unavailable). This does not include intentionally removed information but might include, e.g., redacted text.}

\dsanswer{Some smart-metered electricity consumption data is missing from individual instances in the dataset, primarily due to gaps in the time-series resulting from transmission failures in the data provider's systems. Additionally, there are a few instances of missing survey data, which are attributable to enumeration errors encountered during field data collection---mostly arising from incomplete responses and data entry issues.
}

\dsquestionex{Are relationships between individual instances made explicit (e.g., users' movie ratings, social network links)?}{If so, please describe how these relationships are made explicit.}

\dsanswer{ No explicit relationships exist between individual instances in the dataset, apart from a weak geographic association through the Customer Service Centre (CSC) area to which each household belongs. These CSC identifiers encompass a large area do not imply any direct social, behavioral, or network-based relationships between them.
}

\dsquestionex{Are there recommended data splits (e.g., training, development, validation, testing)?}{If so, please provide a description of these splits, explaining the rationale behind them.}

\dsanswer{The dataset does not have predefined training, validation, or test splits, but users can apply standard techniques based on the research question and their analysis needs.
}

\dsquestionex{Are there any errors, sources of noise, or redundancies in the dataset?}{If so, please provide a description.}

\dsanswer{ There are no known systematic errors, but the dataset may contain potential sources of noise, inconsistencies, or redundancies. Survey data may be affected by recall bias, respondent misunderstandings, or incomplete responses across longitudinal waves. Smart meter data may include gaps in time-series records due to sensor malfunctions, data transmission failures, or storage limitations. Additionally, some readings may contain out-of-range values or anomalies caused by device errors. While preprocessing steps have been applied to improve data quality, users should be aware of these potential issues when conducting analysis.
}

\dsquestionex{Is the dataset self-contained, or does it link to or otherwise rely on external resources (e.g., websites, tweets, other datasets)?}{If it links to or relies on external resources, a) are there guarantees that they will exist, and remain constant, over time; b) are there official archival versions of the complete dataset (i.e., including the external resources as they existed at the time the dataset was created); c) are there any restrictions (e.g., licenses, fees) associated with any of the external resources that might apply to a future user? Please provide descriptions of all external resources and any restrictions associated with them, as well as links or other access points, as appropriate.}

\dsanswer{ The dataset is self-contained and no other external datasets, websites, or resources are linked or required for using the dataset.
}

\dsquestionex{Does the dataset contain data that might be considered confidential (e.g., data that is protected by legal privilege or by doctor-patient confidentiality, data that includes the content of individuals non-public communications)?}{If so, please provide a description.}

\dsanswer{ No, the dataset does not contain confidential data. It has been fully anonymized to ensure privacy and compliance with ethical standards. Each household is assigned a unique identifier that cannot be traced back to real-world individuals or locations. Sensitive information, such as participant names, addresses, and other personally identifiable details, has been either removed or encoded.
}

\dsquestionex{Does the dataset contain data that, if viewed directly, might be offensive, insulting, threatening, or might otherwise cause anxiety?}{If so, please describe why.}

\dsanswer{ No}

\dsquestionex{Does the dataset relate to people?}{If not, you may skip the remaining questions in this section.}

\dsanswer{The dataset primarily does not focus on individuals but includes some (de-identified) information about household inhabitants, only to the extent that it relates to household electricity consumption.
}

\dsquestionex{Does the dataset identify any subpopulations (e.g., by age, gender)?}{If so, please describe how these subpopulations are identified and provide a description of their respective distributions within the dataset.}

\dsanswer{The dataset includes some de-identified information about household inhabitants, primarily in the context of household electricity consumption. While it does not focus on individuals, the longitudinal survey component captures demographic attributes such as age groups, gender, income levels, education, and employment status. These subpopulations are represented as categorical variables in the survey data.

The dataset was designed with a stratified sampling approach to ensure balanced representation across smart-metered and non-smart-metered households, but it is not necessarily representative of Sri Lanka's entire population
}

\dsquestionex{Is it possible to identify individuals (i.e., one or more natural persons), either directly or indirectly (i.e., in combination with other data) from the dataset?}{If so, please describe how.}

\dsanswer{ No, it is impossible to identify individuals directly or indirectly from the dataset as measures have been taken to remove personally identifiable information (PII), and other potential variables for indirect identification.}

\dsquestionex{Does the dataset contain data that might be considered sensitive in any way (e.g., data that reveals racial or ethnic origins, sexual orientations, religious beliefs, political opinions or union memberships, or locations; financial or health data; biometric or genetic data; forms of government identification, such as social security numbers; criminal history)?}{If so, please provide a description.}

\dsanswer{ The dataset contains certain features that could be construed as sensitive, such as detailed physical characteristics of households (e.g., type of dwelling, number of rooms, appliance ownership) and demographic indicators of their members. However, the likelihood of tracing this information back to a specific individual or household is extremely low. All personally identifiable information has been excluded, and the dataset does not contain names, contact details, precise geolocation, or government-issued identifiers.}

\dsquestion{Any other comments?}

\dsanswer{
No
}

\bigskip
\dssectionheader{Collection Process}

\dsquestionex{How was the data associated with each instance acquired?}{Was the data directly observable (e.g., raw text, movie ratings), reported by subjects (e.g., survey responses), or indirectly inferred/derived from other data (e.g., part-of-speech tags, model-based guesses for age or language)? If data was reported by subjects or indirectly inferred/derived from other data, was the data validated/verified? If so, please describe how.}

\dsanswer{ The dataset was acquired through a combination of direct observation, and self-reported survey responses.

\begin{itemize}
    \item \textbf{Directly observed data:} Electricity consumption data was collected from smart meters (at 15-minute and 6-hour intervals) and non-smart meters (monthly aggregate readings), as recorded by Lanka Electricity Company (LECO). Smart meter readings also include voltage, current, and frequency measurements.

    \item \textbf{Self-reported data:} Demographic, geographic, and behavioral information was obtained through a longitudinal survey conducted with household members. This includes details on appliance ownership, usage patterns, energy security, and attitudes toward electricity consumption.
\end{itemize}

To ensure data quality, survey responses were validated through back-checks, supervisor accompaniments, and pilot testing before full deployment. Additionally, smart meter data underwent preprocessing to detect and handle missing values, anomalies, and inconsistencies caused by sensor malfunctions or transmission failures. While efforts were made to improve accuracy, users should consider potential gaps and limitations when analyzing the data.
}

\dsquestionex{What mechanisms or procedures were used to collect the data (e.g., hardware apparatus or sensor, manual human curation, software program, software API)?}{How were these mechanisms or procedures validated?}

\dsanswer{For the survey component, Computer-Assisted Personal Interviewing (CAPI) was used, where trained enumerators conducted face-to-face interviews using tablets to administer the questionnaire and record responses. Survey responses were validated through multiple quality assurance procedures, including supervisor accompaniments, back-checks, and pilot testing prior to full deployment. Enumerators received training to minimize interviewer bias and ensure standardized data collection. The CAPI system also included built-in validation checks to flag and correct errors in real time, reducing the likelihood of inconsistencies or missing responses.

Electricity consumption data was obtained from the Data Center of the Lanka Electricity Company (LECO). This included data from both smart meters and manually read non-smart meters. Preprocessing techniques were applied to this data to identify and address missing values, anomalies, and potential sensor malfunctions, ensuring the reliability of the final dataset.
}

\dsquestion{If the dataset is a sample from a larger set, what was the sampling strategy (e.g., deterministic, probabilistic with specific sampling probabilities)?}

\dsanswer{ The LECO customer database served as the sampling frame.
Households were drawn using stratified random sampling, first by meter type and then by LECO branch area}

\dsquestion{Who was involved in the data collection process (e.g., students, crowdworkers, contractors) and how were they compensated (e.g., how much were crowdworkers paid)?}

\dsanswer{ Trained and experienced enumerators from the project's survey partner, Survey Research Lanka (SRL)  were involved in the collection process
}

\dsquestionex{Over what timeframe was the data collected? Does this timeframe match the creation timeframe of the data associated with the instances (e.g., recent crawl of old news articles)?}{If not, please describe the timeframe in which the data associated with the instances was created.}

\dsanswer{Survey data was collected from 15 November 2023 to 12 Jan 2025 while electricity consumption data was collected from 2023-01-01 to 2024-12-23}

\dsquestionex{Were any ethical review processes conducted (e.g., by an institutional review board)?}{If so, please provide a description of these review processes, including the outcomes, as well as a link or other access point to any supporting documentation.}

\dsanswer{Apart from the Ethics Review Committees established within universities---which are accessible only to researchers affiliated with those institutions---Sri Lanka does not have institutional review boards for obtaining ethics clearance on research projects. However, through its partnership with the DataSEARCH research group at the University of Moratuwa, LIRNEasia was able to secure ethics clearance for this project. This approval letter can be accessed at \href{https://lirneasia2.sharepoint.com/:b:/s/DAPTeam-LACUNA/Ee0alfMaYv1Mnj0ar67_iSQBwAcb3IzIMLzFpRyKB-aFxw?e=nA3fEm}{HERE}
}

\dsquestionex{Does the dataset relate to people?}{If not, you may skip the remaining questions in this section.}

\dsanswer{The dataset does not focus on individuals but includes some (de-identified) information about household inhabitants, only to the extent that it relates to household electricity consumption.
}

\dsquestion{Did you collect the data from the individuals in question directly, or obtain it via third parties or other sources (e.g., websites)?}

\dsanswer{The data was collected through a combination of direct interactions with individuals and third-party sources. Survey data was collected via Survey Research Lanka (SRL) Pvt Ltd, a competitively procured survey partner. While SRL conducted the face-to-face interviews using Computer-Assisted Personal Interviewing (CAPI), our team played an active role in the data collection process by training enumerators, closely supervising pilot surveys, and conducting field accompaniments to ensure data quality and consistency. Electricity consumption data was obtained from Lanka Electricity Company (LECO), which provided de-identified smart meter and non-smart meter readings for participating households.
}

\dsquestionex{Were the individuals in question notified about the data collection?}{If so, please describe (or show with screenshots or other information) how notice was provided, and provide a link or other access point to, or otherwise reproduce, the exact language of the notification itself.}

\dsanswer{Yes, the individuals were notified. Prior to administering the questionnaire, enumerators obtained verbal consent from each participant after clearly explaining the purpose of the study. Participants were also informed that the interview would be audio-recorded, with their permission, to support later quality checks. In addition, they were explicitly notified that their household electricity consumption data would be accessed from the Lanka Electricity Company (LECO) and linked to their survey responses for research purposes. There was a clear mechanism for participants to revoke consent at any point, as explained below.}

\dsquestionex{Did the individuals in question consent to the collection and use of their data?}{If so, please describe (or show with screenshots or other information) how consent was requested and provided, and provide a link or other access point to, or otherwise reproduce, the exact language to which the individuals consented.}

\dsanswer{ Yes, the individuals consented to the collection and use of their data. Before administering the questionnaire, participants were informed about the purpose of the study, how their data would be used, and any potential implications.
}

\dsquestionex{If consent was obtained, were the consenting individuals provided with a mechanism to revoke their consent in the future or for certain uses?}{If so, please provide a description, as well as a link or other access point to the mechanism (if appropriate).}

\dsanswer{Yes, consenting individuals were provided with a mechanism to revoke their consent at any time or for specific uses. Participants were explicitly informed that withdrawing consent would have no consequences. They were given a designated contact point (email and phone) to request the removal of their data. Additionally, they had the option to directly contact Lanka Electricity Company (LECO) through its customer care hotline to address any concerns related to their electricity consumption data.
}

\dsquestionex{Has an analysis of the potential impact of the dataset and its use on data subjects (e.g., a data protection impact analysis) been conducted?}{If so, please provide a description of this analysis, including the outcomes, as well as a link or other access point to any supporting documentation.}

\dsanswer{A comprehensive data protection assessment has not yet been conducted. However, similar to many personal data protection frameworks, Sri Lanka's Personal Data Protection Act, No. 9 of 2022 applies to the processing of personal data. The dataset has been subjected to anonymization techniques, and all reasonable measures have been taken to ensure that no individual can be identified---either directly or indirectly---based on the available data.}

\dsquestion{Any other comments?}

\dsanswer{No.
}

\bigskip
\dssectionheader{Preprocessing/cleaning/labeling}

\dsquestionex{Was any preprocessing/cleaning/ labeling of the data done (e.g., discretization or bucketing, tokenization, part-of-speech tagging, SIFT feature extraction, removal of instances, processing of missing values)?}{If so, please provide a description. If not, you may skip the remainder of the questions in this section.}

\dsanswer{Yes, preprocessing and cleaning were performed to enhance the quality, consistency, and usability of both survey and electricity consumption data. The following steps were undertaken:

\begin{enumerate}
    \item \textbf{Validation of Survey Responses:} Ensuring responses were logical, internally consistent, and aligned with expected patterns through automated checks and manual review.
    \item \textbf{Standardization of Data Formats:} Harmonizing numerical, categorical, and textual variables across all records to maintain consistency.
    \item \textbf{Handling Missing Data:} Identifying and addressing missing survey responses and smart meter gaps using imputation techniques where appropriate.
    \item \textbf{Longitudinal Consistency:} Aligning responses collected across multiple survey waves to ensure coherence over time.
    \item \textbf{De-identification:} Anonymizing personally identifiable information to protect participant privacy while maintaining data utility.
    \item \textbf{Electricity Consumption Data Processing:} Standardizing energy consumption metrics, flagging outliers, and ensuring uniformity in feature representation and time alignment across datasets.
\end{enumerate}
}

\dsquestionex{Was the ``raw'' data saved in addition to the preprocessed/cleaned/labeled data (e.g., to support unanticipated future uses)?}{If so, please provide a link or other access point to the ``raw'' data.}

\dsanswer{No, the raw data was not saved}

\dsquestionex{Is the software used to preprocess/clean/label the instances available?}{If so, please provide a link or other access point.}

\dsanswer{ The dataset was preprocessed and cleaned using a combination of Python and SPSS. The Python scripts used for cleaning and SPSS codes will be made publicly available soon through the project's \href{https://github.com/LIRNEasia/lacuna}{GitHub repository}.}

\dsquestion{Any other comments?}

\dsanswer{No.
}

\bigskip
\dssectionheader{Uses}

\dsquestionex{Has the dataset been used for any tasks already?}{If so, please provide a description.}

\dsanswer{The dataset has been made publicly available through IEEE DataPort, Zenodo, and Kaggle, where it has been downloaded more than 200 times at the time of writing. No peer-reviewed publications using this dataset have been identified to date. Ongoing work by the authors includes a study applying machine learning techniques to identify households with inefficient electricity consumption patterns.}.

\dsquestionex{Is there a repository that links to any or all papers or systems that use the dataset?}{If so, please provide a link or other access point.}

\dsanswer{No such repository exists yet, but known usage of the dataset will be documented in the project Github repository.
}

\dsquestion{What other tasks could the dataset be used for?}

\dsanswer{This dataset can be used for various analytical and predictive tasks. To understand consumer profiles and customer attributes through both unsupervised and supervised machine learning techniques. Load shape analysis can be conducted to study variations in electricity consumption patterns over time. The dataset also supports non-intrusive load monitoring (NILM), allowing for household demand disaggregation to identify individual appliance usage. At the building level, energy prediction models can be developed using statistical or machine learning algorithms, while at the community and regional levels, forecasting energy demand---both in aggregate and during specific periods---can be achieved. Additionally, it facilitates building energy benchmarking to assess relative energy efficiency compared to similar structures. The dataset can further aid in demand management strategies by identifying peak load periods and optimizing energy distribution. Lastly, its longitudinal nature makes it valuable for analyzing behavioral changes in energy consumption over time, driven by factors such as policy shifts, economic conditions, or technological adoption.}

\dsquestionex{Is there anything about the composition of the dataset or the way it was collected and preprocessed/cleaned/labeled that might impact future uses?}{For example, is there anything that a future user might need to know to avoid uses that could result in unfair treatment of individuals or groups (e.g., stereotyping, quality of service issues) or other undesirable harms (e.g., financial harms, legal risks) If so, please provide a description. Is there anything a future user could do to mitigate these undesirable harms?}

\dsanswer{To the best of the authors' knowledge, nothing in the dataset's composition, collection process, or preprocessing (including cleaning and labeling) is expected to impact its future use.
}

\dsquestionex{Are there tasks for which the dataset should not be used?}{If so, please provide a description.}

\dsanswer{The dataset should not be used for tasks requiring national representativeness, as coverage is limited to households within LECO's service area along Sri Lanka's western coastal region. It cannot support appliance-level load disaggregation since consumption is measured only at the household meter level, nor is it suitable for analyses requiring precise income estimates, as only proxies such as expenditure and socio-economic classification are available. Finally, the data are historical and contain transmission gaps, making them inappropriate for real-time operational forecasting or grid control.
}

\dsquestion{Any other comments?}

\dsanswer{No.
}

\bigskip
\dssectionheader{Distribution}

\dsquestionex{Will the dataset be distributed to third parties outside of the entity (e.g., company, institution, organization) on behalf of which the dataset was created?}{If so, please provide a description.}

\dsanswer{Yes, the dataset is publicly available under the Creative Commons Attribution 4.0 International (CC BY 4.0) license. This means that third parties are free to access, use, share, and modify the dataset, provided they give appropriate credit. There are no additional distribution plans beyond making the dataset publicly accessible under this license."
}

\dsquestionex{How will the dataset will be distributed (e.g., tarball on website, API, GitHub)}{Does the dataset have a digital object identifier (DOI)?}

\dsanswer{The dataset is primarily hosted on IEEE DataPort and is assigned the DOI \url{https://dx.doi.org/10.21227/n1dk-q860}. It is also available on Zenodo under the same DOI. Additionally, it is also hosted on Kaggle.
}

\dsquestion{When will the dataset be distributed?}

\dsanswer{The dataset is already publicly available under the DOI provided above.}

\dsquestionex{Will the dataset be distributed under a copyright or other intellectual property (IP) license, and/or under applicable terms of use (ToU)?}{If so, please describe this license and/or ToU, and provide a link or other access point to, or otherwise reproduce, any relevant licensing terms or ToU, as well as any fees associated with these restrictions.}

\dsanswer{The dataset is available under the Creative Commons Attribution 4.0 International (CC BY 4.0) license, allowing third parties are free to access, use, share, and modify the dataset, provided they give appropriate credit. The relevant licensing terms are available at \url{https://creativecommons.org/licenses/by/4.0/deed.en}.}

\dsquestionex{Have any third parties imposed IP-based or other restrictions on the data associated with the instances?}{If so, please describe these restrictions, and provide a link or other access point to, or otherwise reproduce, any relevant licensing terms, as well as any fees associated with these restrictions.}

\dsanswer{There are no imposed IP-based or other restrictions on the data associated with the instances.
}

\dsquestionex{Do any export controls or other regulatory restrictions apply to the dataset or to individual instances?}{If so, please describe these restrictions, and provide a link or other access point to, or otherwise reproduce, any supporting documentation.}

\dsanswer{No.
}

\dsquestion{Any other comments?}

\dsanswer{No.
}

\bigskip
\dssectionheader{Maintenance}

\dsquestion{Who will be supporting/hosting/ maintaining the dataset?}

\dsanswer{LACUNA fund has covered the costs of hosting the dataset on IEEE\textit{DataPort} for a period of 10 years, which we consider a reasonable period for its use. Additionally, the dataset is available on Zenodo and Kaggle---both free hosting platforms---ensuring long-term accessibility unless these services discontinue or change their hosting and access policies.
}

\dsquestion{How can the owner/curator/manager of the dataset be contacted (e.g., email address)?}

\dsanswer{Reach out to info@lirneasia.net via email.
}

\dsquestionex{Is there an erratum?}{If so, please provide a link or other access point.}

\dsanswer{No
}

\dsquestionex{Will the dataset be updated (e.g., to correct labeling errors, add new instances, delete instances)?}{If so, please describe how often, by whom, and how updates will be communicated to users (e.g., mailing list, GitHub)?}

\dsanswer{We do not plan to update the dataset by adding new instances. However, if labeling errors are identified or certain instances need to be removed, we will make the necessary corrections. Any updates will be documented on the dedicated GitHub page of the dataset linked above, and usage instructions detailing this process will be provided on all platforms where the dataset is hosted.
}

\dsquestionex{If the dataset relates to people, are there applicable limits on the retention of the data associated with the instances (e.g., were individuals in question told that their data would be retained for a fixed period of time and then deleted)?}{If so, please describe these limits and explain how they will be enforced.}

\dsanswer{The dataset does not focus on individuals but includes some de-identified information about household inhabitants, only to the extent that it relates to household electricity consumption. There are no restrictions on data retention for these instances.
}

\dsquestionex{Will older versions of the dataset continue to be supported/hosted/maintained?}{If so, please describe how. If not, please describe how its obsolescence will be communicated to users.}

\dsanswer{Aside from the minor updates mentioned above, no new versions of the dataset will be released.
}

\dsquestionex{If others want to extend/augment/ build on/contribute to the dataset, is there a mechanism for them to do so?}{If so, please provide a description. Will these contributions be validated/verified? If so, please describe how. If not, why not? Is there a process for communicating/distributing these contributions to other users? If so, please provide a description.}

\dsanswer{Due to the nature of the dataset---generated through collaboration with an electricity distributor to obtain consumption data and conduct a longitudinal survey with corresponding households---extending it is not a straightforward process. However, researchers or practitioners interested in augmenting, building on, or contributing to the dataset are encouraged to reach out via the provided email. This information is also available in the \href{https://github.com/LIRNEasia/lacuna}{GitHub repository}.
}

\dsquestion{Any other comments?}

\dsanswer{No
}

\end{document}